# Distributed Computation of Large-scale Graph Problems

Hartmut Klauck*    Danupon Nanongkai†    Gopal Pandurangan‡    Peter Robinson§

**Abstract**

Motivated by the increasing need for fast distributed processing of large-scale graphs such as the Web graph and various social networks, we study a number of fundamental graph problems in the message-passing model, where we have $k$ machines that jointly perform a computation on an arbitrary $n$-node (typically, $n \gg k$) input graph. The graph is assumed to be *randomly* partitioned among the $k \geq 2$ machines (a common implementation in many real world systems). The communication is point-to-point, and the goal is to minimize the time complexity, i.e., the number of communication rounds, of solving various fundamental graph problems.

We present lower bounds that quantify the fundamental time limitations of distributively solving graph problems. We first show a lower bound of $\Omega(n/k)$ rounds for computing a spanning tree (ST) of the input graph. This result also implies the same bound for other fundamental problems such as computing a minimum spanning tree (MST), breadth-first tree (BFS), and shortest paths tree (SPT). We also show an $\Omega(n/k^2)$ lower bound for connectivity, ST verification and other related problems. Our lower bounds develop and use new bounds in *random-partition* communication complexity.

To complement our lower bounds, we also give algorithms for various fundamental graph problems, e.g., PageRank, MST, connectivity, ST verification, shortest paths, cuts, spanners, covering problems, densest subgraph, subgraph isomorphism, finding triangles, etc. We show that problems such as PageRank, MST, connectivity, and graph covering can be solved in $\tilde{O}(n/k)$ time (the notation $\tilde{O}$ hides polylog($n$) factors and an additive polylog($n$) term); this shows that one can achieve almost *linear* (in $k$) speedup, whereas for shortest paths, we present algorithms that run in $\tilde{O}(n/\sqrt{k})$ time (for $(1+\epsilon)$-factor approximation) and in $\tilde{O}(n/k)$ time (for $O(\log n)$-factor approximation) respectively.

Our results are a step towards understanding the complexity of distributively solving large-scale graph problems.

## 1 Introduction

The emergence of "Big Data" over the last decade or so has led to new computing platforms for distributed processing of large-scale data, exemplified by MapReduce [12] and more recently systems such as Pregel [29] and Giraph[34]. In these platforms, the data — which is simply too large to fit into a single machine — is distributed across a group of machines that are connected via a communication network and the machines jointly process the data in a distributed fashion. The focus of this paper is on distributed processing of large-scale *graphs* which is increasingly becoming important with the rise of massive graphs such as the Web graph, social networks, biological networks, and other graph-structured data and the consequent need for fast graph algorithms on such large-scale graph data. Indeed, there has been a recent proliferation of systems designed specifically for large-scale graph processing, e.g., Pregel [29], Giraph [34], GraphLab [26], and GPS [1]. MapReduce (developed at Google [12]) has become a very successful distributed computing platform for a wide variety of large-scale computing applications and also has been used for processing graphs [25]. However, as pointed out by the developers of Pregel (which was also developed at Google), MapReduce may sometimes be ill-suited for implementing graph algorithms; it can lead to "suboptimal performance and usability issues" [29]. On the other hand, they mention that graph algorithms seem better suited to a *message-passing* distributed computing model [33, 28] and this is the main design principle [29] behind Pregel (and other systems that followed it such as Giraph [34] and GPS [1]). While there is a rich theory for the message-passing distributed computing model [33, 28], such a theory is still in its infancy for distributed graph processing systems.

*Division of Mathematical Sciences, Nanyang Technological University, Singapore 637371 & Centre for Quantum Technologies, Singapore 117543. E-mail: `hklauck@gmail.com`. This work is funded by the Singapore Ministry of Education (partly through the Academic Research Fund Tier 3 MOE2012-T3-1-009) and by the Singapore National Research Foundation.

†KTH Royal Institute of Technology, Sweden, and University of Vienna, Austria E-mail: `danupon@gmail.com`. Work done while at ICERM, Brown University, USA, and Nanyang Technological University, Singapore.

‡Department of Computer Science, University of Houston, Houston, TX 77204, USA. E-mail: `gopalpandurangan@gmail.com`. Work done while at Division of Mathematical Sciences, Nanyang Technological University, Singapore 637371 & Department of Computer Science, Brown University, Providence, RI 02912, USA. Supported in part by the following research grants: Nanyang Technological University grant M58110000, Singapore Ministry of Education (MOE) Academic Research Fund (AcRF) Tier 2 grant MOE2010-T2-2-082, Singapore MOE AcRF Tier 1 grant MOE2012-T1-001-094, and a grant from the US-Israel Binational Science Foundation (BSF).

§Department of Computer Science, National University of Singapore. E-mail: `robinson@comp.nus.edu.sg`. Research supported by the grant MOE2011-T2-2-042 "Fault-tolerant Communication Complexity in Wireless Networks" from the Singapore MoE AcRF-2.

In this work, our goal is to investigate a theory for large-scale graph computation based on a distributed message-passing model. A fundamental issue that we would like to investigate is the amount of "speedup" possible in such a model vis-a-vis the number of machines used: more precisely, if we use $k$ machines, does the run time scale linearly (or even super-linearly) in $k$? And what are the fundamental time bounds for various graph problems?

**1.1 Model** We consider a *network* of $k > 1$ (distinct) *machines* $N = \{p_1, \ldots, p_k\}$ that are pairwise interconnected by bidirectional point-to-point communication *links* — henceforth called the *k-machine* model[1]. Each machine executes an instance of a distributed algorithm $A$. The computation advances in *synchronous* rounds where, in each round, machines can exchange messages over their communication links. Each link is assumed to have a bandwidth of $W$, i.e., $W$ bits can be transmitted over the link in one round. (In stating our time bounds, for convenience, we will assume that bandwidth $W = 1$; in any case, it is easy to rewrite our upper bounds to scale in terms of parameter $W$ — cf. Theorem 4.1). (Note that machines have no other means of communication and do not share any memory.) There is an alternate — but equivalent — way to view our communication restriction: instead of putting bandwidth restriction on the links, we can put a restriction on the amount of information that each *machine* can communicate (i.e. send/receive) in a round. The results that we obtain in the bandwidth-restricted model will also apply to the latter model (cf. Section 4.2). Local computation within a machine is considered free, while communicating messages between the machines is the costly operation[2].

We are interested in solving graph problems where we are given an *input graph* $G$ of $n$ *vertices* (assume that each vertex has a unique label) and $m$ *edges* from some *input domain* $\mathcal{G}$. To avoid trivialities, we will assume that $n \geq k$ (typically $n \gg k$). Unless otherwise stated, we will consider $G$ to be undirected, although all our results can be made to apply in a straightforward fashion to directed graphs as well. Initially, the entire graph $G$ is not known by a single machine, but rather partitioned among the $k$ machines in a *"balanced"* fashion, i.e., the nodes and/or edges of $G$ must be partitioned approximately evenly among the machines. We will assume a *vertex-partition* model, where vertices (and their incident edges) are partitioned across machines. One type of partition that we will assume throughout is the *random (vertex)* partition, i.e., vertices (and its incident edges) of the input graph are assigned randomly to machines. (This is the typical way that many real systems (e.g., Pregel) partition the input graph among the machines; it is simple and and easy to accomplish, e.g., via hashing [3].) Our upper bounds will also hold (with slight modifications) without this assumption; only a *"balanced"* partition of the input graph among the machines is needed. On the other hand, our lower bounds apply even under random partitioning, hence they apply to worst-case partition as well. It can be shown that (cf. Lemma 4.1) a random partition gives rises to an (approximately) balanced partition.

Formally, in the *random vertex partition (RVP)* model, each vertex of $G$ is assigned independently and randomly to one of the $k$ machines. If a vertex $v$ is assigned to machine $p_i$ we call $p_i$ the *home* machine of $v$. Note that when a vertex is assigned to a machine, *all its incident edges* are assigned to that machine as well; i.e., the home machine will know the labels of neighbors of that vertex as well as the identity of the home machines of the neighboring vertices. A convenient way to implement the above assignment is via *hashing*: each vertex (label) is hashed to one of the $k$ machines. Hence, if a machine knows a vertex label, it also knows where it is hashed to.

Depending on the problem $\mathcal{P}$, the vertices and/or edges of $G$ have labels chosen from a set of polynomial (in $n$) size. Eventually, each *machine* $p_i$ ($1 \leq i \leq k$) must (irrevocably) set a designated local output variable $o_i$ (which will depend on the set of vertices assigned to machine $p_i$) and the *output configuration* $o = \langle o_1, \ldots, o_k \rangle$ must satisfy certain feasibility conditions w.r.t. problem $\mathcal{P}$. For example, when considering the minimum spanning tree (MST) problem, each $o_i$ corresponds to a set of edges (which will be a subset of edges incident on vertices mapped to machine) $p_i$ and the edges in the union of the sets $o_i$ must form an MST of the input graph $G$; in other words, each machine $p_i$ will know all the MST edges incident on vertices mapped to $p_i$. (Note that this is a natural generalization of the analogous assumption in the standard distributed message passing model, where each vertex knows which of its incident edges belong to the MST [33].) We say that *algorithm A solves problem $\mathcal{P}$* if $A$ maps each $G \in \mathcal{G}$ to an output configuration that is feasible for $\mathcal{P}$. The *time complexity of A* is the maximum number of rounds until termination, over all graphs in $\mathcal{G}$. In stating our time bounds, for convenience, we will assume that bandwidth $W = 1$; in any case, it is easy to rewrite our upper bounds to scale in terms of parameter $W$ (cf. Theorem 4.1).

---

[1]Our results can also be generalized to work if the communication network is a sparse topology as well if one assumes an underlying routing mechanism; details are omitted here.

[2]This assumption is reasonable in the context of large-scale data, e.g., it has been made in the context of theoretical analysis of MapReduce, see e.g., [35] for a justification. Indeed, typically in practice, even assuming the links have a bandwidth of order of gigabytes of data per second, the amount of data that have been to be communicated can be in order of tera or peta bytes which generally dominates the overall computation cost [35].

[3]Partitioning based on the structure of the graph — with the goal of minimizing the amount of communication between the machines — is non-trivial; finding such a "good" partition itself might be prohibitively expensive and can be problem dependent. Some papers address this issue, see e.g., [40, 3, 39].

**Notation.** For any $0 \leq \epsilon \leq 1$, we say that a protocol has *$\epsilon$-error* if, for any input graph $G$, it outputs the correct answer with probability at least $1 - \epsilon$, where the probability is over the random partition and the random bit strings used by the algorithm (in case it is randomized).

For any $n > 0$ and function $T(n)$, we say that an algorithm $\mathcal{A}$ *terminates in $O(T(n))$ rounds* if, for any $n$-node graph $G$, $\mathcal{A}$ always terminate in $O(T(n))$ rounds, regardless of the choice of the (random) input partition. For any $n$ and problem $\mathcal{P}$ on $n$ node graphs, we let the *time complexity of solving $\mathcal{P}$ with $\epsilon$ error probability* in the $k$-machine model, denoted by $\mathcal{T}_\epsilon^k(\mathcal{P})$, be the minimum $T(n)$ such that there exists an $\epsilon$-error protocol that solves $\mathcal{P}$ and terminates in $T(n)$ rounds. For any $0 \leq \epsilon \leq 1$, graph problem $\mathcal{P}$ and function $T: \mathbb{Z}_+ \to \mathbb{Z}_+$, we say that $\mathcal{T}_\epsilon^k(\mathcal{P}) = O(T(n))$ if there exists integer $n_0$ and $c$ such that for all $n \geq n_0$, $\mathcal{T}_\epsilon^k(\mathcal{P}) \leq cT(n)$. Similarly, we say that $\mathcal{T}_\epsilon^k(\mathcal{P}) = \Omega(T(n))$ if there exists integer $n_0$ and real $c$ such that for all $n \geq n_0$, $\mathcal{T}_\epsilon^k(\mathcal{P}) \geq cT(n)$. For our upper bounds, we will usually use $\epsilon = 1/n$, which will imply high probability algorithms, i.e., succeeding with probability at least $1 - 1/n$. In this case, we will sometimes just omit $\epsilon$ and simply say the time bound applies "with high probability". We use $\Delta$ to denote the maximum degree of any node in the input graph, and $D$ for the diameter of the input graph.

**1.2  Our Results and Techniques** Our main goal is to investigate the *time* complexity, i.e., the number of distributed "rounds", for solving various fundamental graph problems. The time complexity not only captures the (potential) speed up possible for a problem, but it also implicitly captures the communication cost of the algorithm as well, since links can transmit only a limited amount of bits per round; equivalently, we can view our model where instead of links, *machines* can send/receive only a limited amount of bits per round. We develop techniques to obtain non-trivial lower and upper bounds on the time complexity of various graph problems.

**Lower Bounds.** Our lower bounds quantify the fundamental time limitations of distributively solving graph problems. They apply essentially to distributed data computations in all point-to-point communication models, since they apply even to a synchronous complete network model where the graph is partitioned *randomly* (unlike some previous results, e.g., [41], which apply only under some worst-case partition).

We first give a tight lower bound on the complexity of computing a spanning tree (cf. Section 2). The proof shows that $\Omega(n/k)$ rounds of communication are needed even for unweighted and undirected graphs of diameter 2, and even for sparse graphs. We give an information theoretic argument for this result. This result also implies the same bound for other fundamental problems such as computing a minimum spanning tree, breadth-first tree, and shortest paths tree. This bound shows that one cannot hope to obtain a run time that scales (asymptotically) faster than $1/k$. This result, in conjunction with our upper bound of $\tilde{O}(n/k)$ for computing a MST, shows that this lower bound is essentially tight.

We then show an $\Omega(n/k^2)$ lower bound for connectivity, spanning tree (ST) verification and other related verification problems (cf. Section 3). To analyze the complexity of verification problems we give reductions from problems in the 2-player communication complexity model using random partitions of the input variables. As opposed to the standard fixed partition model here all input bits are *randomly* assigned to Alice and Bob. We give a tight lower bound for randomized protocols for the well-studied *disjointness* problem in this setting. In particular, we show a lower bound on the randomized *average partition* communication complexity of the disjointness problem which might be of independent interest. Random partition communication complexity has also been studied by Chakrabarti et al. [8], but their results apply to the promise disjointness problem in the multiparty number in hand model for a sufficiently large number of players only. In our proof we apply the rectangle based arguments of Razborov [36], but we need to take care of several issues arising. A core ingredient of Razborov's proof is the conditioning that turns the input distribution into a product distribution. With randomly assigned inputs we need to recover the necessary product properties by conditioning over badly assigned input bits. Even when doing so the size of the sets in the input are no longer exactly as in Razborov's proof. Furthermore, there is a large probability that the set intersection is visible to a single player right from the start, but with a small enough error probability the communication still needs to be large.

**Algorithms and Upper Bounds.** We introduce techniques to obtain fast graph algorithms in the $k$-machine model (cf. Section 1.1). We first present a general result, called the *Conversion Theorem* (cf. Theorem 4.1) that, given a graph problem $\mathcal{P}$, shows how fast algorithms for solving $\mathcal{P}$ in the $k$-machine model can be designed by leveraging distributed algorithms for $\mathcal{P}$ in the standard CONGEST message-passing distributed computing model (see e.g., [33]). We note that fast distributed algorithms in the standard model *do not* directly imply fast algorithms in the $k$-machine model. To achieve this, we consider distributed algorithms in an intermediate *clique* model (cf. Section 4.2) and then show two ways — parts (a) and (b) respectively of the Conversion Theorem — to efficiently convert algorithms in the clique model to the $k$-machine model. Part (b) applies to converting distributed algorithms (in the clique model) that only uses broadcast, while part (a) applies to any algorithm. Part (a) will sometimes give better time bounds compared to part (b) and vice versa — this depends on the problem at hand and the type of distributed algorithm considered, as well as on the graph parameters. (The latter can be especially useful in applications where we might have some information on the graph parameters/topology as explained below.) Using this theorem, we design algorithms for various fundamental graph problems, e.g., PageRank, minimum spanning tree (MST), connectivity, spanning tree (ST) verification, shortest paths, cuts, spanners, covering problems, densest subgraph, subgraph isomorphism, Triangle finding (cf. Table 1). We show that problems such as PageRank, MST, and connectivity, graph covering etc. can be solved in $\tilde{O}(n/k)$ time; this shows that one can achieve almost *linear* (in $k$) speedup. For graph connectivity, BFS tree construction, and ST verification, we show $\tilde{O}(\min(n/k, m/k^2 + D\Delta/k))$ bound — note that the second part of the above bound may be better in some cases, e.g., if the graph is sparse (i.e., $m = O(n)$) and $D$ and $\Delta$ are small (e.g., bounded by $O(\log n)$) — then we get a bound of $\tilde{O}(n/k^2)$. For single-source shortest paths, another classic and important problem, we show a bound of $\tilde{O}(n/\sqrt{k})$ for a $(1+\epsilon)$-factor approximation and a bound of $\tilde{O}(n/k)$ for $O(\log n)$-factor approximation. We note that if one wants to compute *exact* shortest paths, this might take significantly longer (e.g., using Bellman-Ford — cf. Section 4.3). For graph covering problems such a Maximal Independent Set (MIS) and (approximate) Minimum Vertex cover (MVC), we show a bound of $\tilde{O}(\min(n/k, m/k^2 + \Delta/k))$; note that this implies a bound of $\tilde{O}(n/k^2)$ for *(constant) bounded degree* graphs, i.e., we can get a speed up that scales superlinearly in $k$.

We finally note that our results also directly apply to an alternate (but equivalent) model, where instead of having a restriction on the number of bits individual links can transmit in a round, we restrict the number of bits a machine can send/receive (in total) per round (cf. Section 4.2).

**1.3 Related Work** The theoretical study of (large-scale) graph processing in distributed systems is relatively recent. Several works have been devoted to developing MapReduce graph algorithms (e.g., see [25, 35] and the references therein). There also have been several recent theoretical papers analyzing MapReduce algorithms in general, including Mapreduce graph algorithms see e.g., [24, 35, 19] and the references therein. We note that the flavor of theory developed for MapReduce is quite different compared to the distributed complexity results of this paper. Minimizing communication (which leads in turn to minimizing the number of communication rounds) is a key motivation in MapReduce algorithms (e.g., see [35]); however this is generally achieved by making sure that the data is made small enough quickly (in a small number of MapReduce rounds) to fit into the *memory* of a single machine. An example of this idea is the filtering technique of [24] applied to graph problems. The main idea behind filtering is to reduce the size of the input in a distributed fashion so that the resulting, much smaller, problem instance can be solved on a single machine. Filtering allows for a tradeoff between the number of rounds and the available memory. Specifically, the work of [24] shows that for graphs with at most $n^{1+c}$ edges and machines with memory at least $n^{1+\varepsilon}$ will require $O(c/\epsilon)$ (MapReduce) rounds.

The work that is closest in spirit to ours is the recent work of [41]. The above work considers a number of basic statistical and graph problems in the message-passing model (where the data is distributed across a set of machines) and analyzes their communication complexity — which denotes the total number of bits exchanged in all messages across the machines during a computation. Their main result is that *exact* computation of many statistical and graph problems in the distributed setting is very expensive, and often one cannot do better than simply having all machines send

| Problem | Upper Bound | Lower Bound |
| --- | --- | --- |
| Minimum Spanning Tree (MST) | $\tilde{O}(n/k)$ | $\tilde{\Omega}(n/k)^*$ |
| Connectivity, Spanning Tree Verification (Conn,ST) | $\tilde{O}(\min(n/k, m/k^2 + D\lceil\Delta/k\rceil))$ | $\tilde{\Omega}(n/k^2)$ |
| Breadth First Search Tree (BFS) | $\tilde{O}(\min(n/k + D, m/k^2 + D\lceil\Delta/k\rceil))$ | $\tilde{\Omega}(n/k)$ |
| Single-Source Shortest-Paths Distances (SSSP) | $\tilde{O}(n/\sqrt{k})^\dagger$, $\tilde{O}(n/k)^\$$ | |
| Single-Source Shortest-Paths Tree (SPT) | $\tilde{O}(n/\sqrt{k})^\dagger$, $\tilde{O}(n/k)^\$$ | $\tilde{\Omega}(n/k)^*$ |
| All-Pairs Shortest-Paths Distances (APSP) | $\tilde{O}(n\sqrt{n}/k)^\#$, $\tilde{O}(n/k)^\$$ | |
| PageRank with reset prob. $\gamma$ (PageRank) | $\tilde{O}(n/\gamma k)$ | |
| Graph Covering Problems (MIS, MVC) | $\tilde{O}(\min(n/k, m/k^2 + \Delta/k))$ | |
| Maximal Ind. Set on Hypergraphs (HMIS) | $\tilde{O}(n/k + k)$ | |
| $(2\delta - 1)$-Spanner (Spanner) ($\delta \in O(\log n)$) | $\tilde{O}(n/k)$ | |
| Densest Subgraph (DSGraph) | $\tilde{O}(n/k)$ (for $(2+\epsilon)$-approx.) | |
| Triangle Verification (Tri) | $\tilde{O}(\min(n\Delta^2/k^2 + \Delta\lceil\Delta/k\rceil, n^{7/3}/k^2 + n^{4/3}/k))$ | |
| Subgraph Isomorphism (SubIso) ($d$-vertex subgraph) | $\tilde{O}(n^{2+(d-2)/d}/k^2 + n^{1+(d-2)/d}/k)$ | |

$\dagger$ $(1 + \epsilon)$-approximation.    $\#$ $(2 + \epsilon)$-approximation.    $\$$ $O(\log n)$-approximation.    $*$ For any approx. ratio.

Table 1: Complexity bounds in the $k$-machine model for an $n$-node input graph with $m$ edges, max degree $\Delta$, and diameter $D$. $\epsilon > 0$ is any small constant. The notation $\tilde{O}$ hides polylog($n$) factors and an additive polylog($n$) term. For clarity of presentation, we assume a bandwidth of $\Theta(\log n)$ bits.

their data to a centralized server. The graph problems considered are computing the degree of a vertex, testing cycle-freeness, testing connectivity, computing the number of connected components, testing bipartiteness, and testing triangle-freeness. The strong lower bounds shown for these assume a *worst-case* distribution of the input (unlike ours, which assumes a random distribution). They posit that in order to obtain protocols that are communication-efficient, one has to allow approximation, or investigate the distribution or layout of the data sets and leave these as open problems for future. Our work, on the other hand, addresses time (round) complexity (this is different from the notion of round complexity defined in [41]) and shows that non-trivial speed up is possible for many graph problems. As posited above, for some problems such as shortest paths and densest subgraph etc., our model assumes a *random partition* of the input graph and also allows *approximation* to get good speedup, while for problems such as MST we get good speedups for exact algorithms as well. For spanning tree problems we show tight lower bounds as well.

The $k$-machine model is closely related to the well-studied (standard) message-passing CONGEST model [33], in particular to the CONGEST clique model (cf. Section 4). The main difference is that while many vertices of the input graph are mapped to the same machine in the $k$-machine model, in the standard model each vertex corresponds to a dedicated machine. More "local knowledge" is available per vertex (since it can access for free information about other vertices in the same machine) in the $k$-machine model compared to the standard model. On the other hand, all nodes assigned to a machine have to communicate through the links incident on this machine, which can limit the bandwidth (unlike the standard model where each vertex has a dedicated processor). These differences manifest in the time complexity — certain problems have a faster time complexity in one model compared to the other (cf. Section 4). In particular, the fastest known distributed algorithm in the standard model for a given problem, may not give rise to the fastest algorithm in the $k$-machine model. Furthermore, the techniques for showing the complexity bounds (both upper and lower) in the $k$-machine model are different compared to the standard model. The recently developed communication complexity techniques (see e.g, [37, 32, 15]) used to prove lower bounds in the standard CONGEST model are not applicable here.

## 2 Tight Lower Bounds for Spanning Tree Computation

In this section, we show that any $\varepsilon$-error algorithm takes $\Omega(n/k)$ rounds to *compute a spanning tree* (STC). Recall that in the STC problem, for every edge $uv$, the home machines of $u$ and $v$ must know whether $uv$ is in the ST or not.

THEOREM 2.1. (LOWER BOUND FOR STC) *Every public-coin $\epsilon$-error randomized protocol on a $k$-machine network that computes a spanning tree of an $n$-node input graph has an expected round complexity of*

$\Omega\left(\frac{n}{k}\right)$. More specifically, there exists a constant $\epsilon > 0$ such that, for any $k \geq 3$ and large enough $n$, $\mathcal{T}_\epsilon^k(\mathsf{STC}), \mathcal{T}_\epsilon^k(\mathsf{MST}), \mathcal{T}_\epsilon^k(\mathsf{BFS}), \mathcal{T}_\epsilon^k(\mathsf{SPT}) \in \Omega\left(\frac{n}{k}\right)$.

*Proof.* We first show the theorem for $\mathcal{T}_\epsilon^k(\mathsf{STC})$ using an information theoretic argument. Assume for a contradiction that there exists a distributed algorithm in the $k$-machine model, denoted by $\mathcal{R}$, that violates Theorem 2.1. In other words, $\mathcal{R}$ solves STC correctly with probability at least $1 - \epsilon$ and always terminates in $\frac{\delta n}{k}$ rounds, for some $\delta \in o(1)$. We will show that the information flow to at least one machine must be large.

**Graph $G_b(X, Y)$.** Let $b = n - 2$. For any $X, Y \subseteq [b]$, we construct the following graph, denoted by $G_b(X, Y)$. The vertices (and adjacent edges of $G_b(X, Y)$) will be assigned to a random machine.

$G_b(X, Y)$ consists of $b + 2$ nodes, denoted by $v_1, \ldots, v_b$ and $u, w$. For each $1 \leq i \leq b$, we add edge $uv_i$ to $G_b(X, Y)$ if $i \in X$, and we add edge $v_i w$ to $G_b(X, Y)$ if $i \in Y$.

The random strings $X, Y$ will be drawn from a distribution that will ensure that $G_b(X, Y)$ is connected. Furthermore the graph will contain roughly $4b/3$ edges with high probability ($2b/3$ adjacent to $u, w$ respectively), hence roughly $b/3$ edges must be removed to obtain a spanning tree.

To produce the correct output in $\mathcal{R}$ the machine that receives $u$ must know which edges $uv_i$ are included in the spanning tree, and similarly the machine that receives $w$ must know which edges $v_i w$ are included in the spanning tree. Since $X, Y$ are encoded in the graph, the machine $p_1$ who receives $u$ knows $X$ via the edges $uv_i$ at the start, but has initially limited information about $Y$, unless it also receives $w$, which happens with probability $1/k$. With probability $1 - 1/k$ the information about $Y$ initially held by $p_1$ comes from the set of vertices $v_i$ held by $p_1$ which is of size $\approx b/k$ with high probability, giving $p_1$ at most $\approx b/k$ bits of information about $Y$. Hence all information needed to decide which edges to not include must come into $p_1$ via communication with the other $k-1$ machines. We show that this communication is $\Omega(b)$, if $p_1$ outputs at most $b/2$ edges as part of the spanning tree (and a symmetric argument holds for the machine $p_2$ that holds $w$). Hence the number of rounds is $\Omega(b/k) = \Omega(n/k)$.

In order to give a clean condition on which edges are necessary for a spanning tree we use the following distribution on $X, Y$. $X, Y$ (viewed as characteristic vectors) are chosen uniformly from $\{0, 1\}^b \times \{0, 1\}^b$ under the condition that for every $i \in [b]$ we have $X_i + Y_i \geq 1$. Hence there are exactly $3^b$ possible values for $X, Y$. The following simple observation is crucial to our proof.

OBSERVATION 1. *For any $X, Y$ chosen as described above, and all $1 \leq i \leq b$ such that $X_i = Y_i = 1$ exactly one of the edges $uv_i$ or $v_i w$ must be part of any spanning tree, except for exactly one such $i$, for which both edges are in the spanning tree. For all other $i$ the one edge $uv_i$ or $v_i w$ that is in the graph must also be in the spanning tree.*

Since all edges are adjacent to either $u$ or $w$, and any spanning tree must have $b + 1$ edges, one of $p_1$ or $p_2$ must output at most $b/2$ edges. Before the first round of communication the entropy $H(Y|X)$ is $2b/3$ by the following calculation:

$$\begin{aligned}
H(Y|X) &= \sum_x \Pr(X = x) \cdot H(Y|X = x) \\
&= 3^{-b} \sum_{\ell=0}^{b} \binom{b}{\ell} 2^\ell \cdot \log 2^\ell \\
&= 3^{-b} b \sum_{\ell=0}^{b-1} \binom{b-1}{\ell} 2^{\ell+1} \\
&= 2b/3.
\end{aligned}$$

Besides $X$ the machine $p_1$ also knows some vertices $v_i$ and their edges, giving it access to some bits of $Y$. It is easy to see via the Chernoff bound that with very high probability $p_1$ knows at most $(1 + \zeta)b/k$ bits of $Y$ for $\zeta = 0.01$, lowering the conditional entropy of $Y$ given those bits to no less than $2b/3 - (1 + \zeta)b/k$. The event where $p_1$ knows more cannot influence the entropy by more than $2^{-\zeta^2 b/(3k)} \cdot b = o(1)$ (for $b$ large enough). Hence the entropy of $Y$ given the initial information of $p_1$, which we denote by a random variable $A$, is $H(Y|A) \geq 2b/3 - (1 + \zeta)b/k - o(1)$.

Assume that $p_1$ outputs at most $b/2$ edges. This event or the corresponding event for $p_2$ must happen with probability $1 - \epsilon$ assuming failure probability $\epsilon$. Conditioned on the event that $p_1$ outputs $b/2$ or fewer edges we can estimate the entropy $H(Y|A, T_0)$ for the random variable $T_0$ containing the transcript of all messages to $p_1$. $H(Y|A, T_0) \leq H(Y|X, E)$ where $E$ is the random variables of edges in the output of $p_1$ (we use that $X$ and $E$ can be computed from $A, T_0$). Given that $X = x$ there are $|x|$ possible edges for $E$. With probability $1 - o(1)$ we have $|y| < 2b/3 + \zeta b$.

For at most $b/2$ edges in $E$ there are at most

$$\sum_{\ell < b/6 + \zeta b} \binom{b/2}{\ell} \leq b \cdot \binom{b/2}{b/6 + \zeta b}$$

possibilities for $Y = y$ such that $|y| \leq 2b/3 + \zeta b$. Hence

we can estimate the remaining entropy of $Y$ as follows:

$$\begin{aligned} H(Y|X,E) &\leq \Pr(|Y| < 2b/3 + \zeta b)(\log \binom{b/2}{b/6+\zeta b} \\ &\quad + \log b) + o(1) \\ &\leq H(1/3 + 2\zeta)b/2 + o(b) \end{aligned}$$

for the binary entropy function $H$. For $k \geq 7$ we can simply use the upper bound $b/2$ for the above quantity and conclude that $I(T_0 : Y|A) = H(Y|A) - H(Y|A,T_0) \geq 2b/3 - (1+\zeta)b/k - o(1) - b/2 \geq \Omega(b)$, and hence $p_1$ must have received messages of length $\Omega(b)$. This happens with some probability $\gamma$ for $p_1$ and with probability $(1-\epsilon) - \gamma$ for $p_2$. Hence $|T_0| + |T_1| \geq \Omega(b)$.

The above analysis is under the assumption that different machines hold $u, w$, which happens with probability $1 - 1/k$. Without this assumption the information flow must be at least $(1 - 1/k) \cdot \Omega(b)$.

For $k < 7$ we need to make a more careful analysis. For instance, $p_1$ actually gets only around $2b/(3k)$ bits of information from knowing $b/k$ bits of $Y$, and the estimate on $H(Y|A,E)$ needs to be made more precise. We skip the details. This completes the proof of $\mathcal{T}_\epsilon^k(\mathsf{STC}) = \Omega\left(\frac{n}{k}\right)$.

To see that this also implies $\mathcal{T}_\epsilon^k(\mathsf{MST}), \mathcal{T}_\epsilon^k(\mathsf{BFS}), \mathcal{T}_\epsilon^k(\mathsf{SPT}) \in \Omega\left(\frac{n}{k}\right)$, it is sufficient to observe that any BFS tree (resp. MST, and SPT, for any approximation ratio) is also a spanning tree.

## 3 Lower Bounds for Verification Problems

In this section, we show lower bounds for the spanning tree (ST) and connectivity (Conn) verification problems. An algorithm solves the Conn verification problem in our model if the machines output 1 if and only if the input graph $G$ is connected; the ST problem is defined similarly. We note that our lower bounds hold even when we allow *shared randomness*, i.e. even when all machines can read the same random string. For a problem $\mathcal{P}$ where a two-sided error is possible, we define the *time complexity of solving $\mathcal{P}$ with $(\epsilon_0, \epsilon_1)$ error probabilities*, denoted by $\mathcal{T}_{\epsilon_0,\epsilon_1}^k(\mathcal{P})$, be the minimum $T(n)$ such that there exists a protocol that solves $\mathcal{P}$, terminates in $T(n)$ rounds, and errs on 0-input with probability at most $\epsilon_0$ and errs on 1-input with probability at most $\epsilon_1$.

THEOREM 3.1. (ST VERIFICATION AND Conn) *There exists a constant $\epsilon > 0$ such that, for any $k \geq 2$ and large enough $n$, $\mathcal{T}_{\epsilon,0}^k(\mathsf{ST}) = \tilde{\Omega}\left(\frac{n}{k^2}\right)$ and $\mathcal{T}_{\epsilon,\epsilon}^k(\mathsf{Conn}) = \tilde{\Omega}\left(\frac{n}{k^2}\right)$. In other words, there is no public-coin $(\epsilon, 0)$-error randomized protocol on a $k$-machine model that, on any $n$-node input graph, solves ST correctly in* $o(\frac{n}{k^2 \log n})$ *expected rounds, and no $(\epsilon, \epsilon)$-error protocol that solves Conn correctly in $o(\frac{n}{k^2 \log n})$ rounds.*

To prove Theorem 3.1, we introduce a new model called *random-partition (two-party) communication complexity* and prove some lower bounds in this model. This is done in Section 3.1. We then use these lower bounds to show lower bounds for the $k$-machine model in Sections 3.1.1 and 3.1.2.

**3.1 Random-Partition Communication Complexity** We first recall the standard communication complexity model, which we will call *worst-partition* model, to distinguish it from our random-partition model. (For a comprehensive review of the subject, we refer the reader to [22].)

In the **worst-partition** communication complexity model, there are two players called Alice and Bob. Each player receives a $b$-bit binary string, for some integer $b \geq 1$. We denote the string received by Alice and Bob by $x$ and $y$ respectively. Together, they both want to compute $f(x, y)$ for a Boolean function $f : \{0,1\}^b \times \{0,1\}^b \to \{0,1\}$. At the end of the process, we want both Alice and Bob to know the value of $f(x, y)$. We are interested in the number of bits exchanged between Alice and Bob in order to compute $f$. We say that a protocol $\mathcal{R}$ has *complexity $t$* if it always uses at most $t$ bits in total[4]. For any function $f$, the *worst-partition communication complexity* for computing $f$ with $(\epsilon_0, \epsilon_1)$-error, denoted by $R_{\epsilon_0,\epsilon_1}^{cc-pub}(f)$, is the minimum $t$ such that there is an $(\epsilon_0, \epsilon_1)$-error protocol with complexity $t$. (Note that a protocol is $(\epsilon_0, \epsilon_1)$-error if it outputs 0 with probability at least $1 - \epsilon_0$ when $f(x,y) = 0$ and outputs 1 with probability at least $1 - \epsilon_1$ when $f(x,y) = 1$.)

The **random-partition** model is slightly different from the worst-partition model in that, instead of giving every bit of $x$ to Alice and every bit of $y$ to Bob, each of these bits are sent to one of the players *randomly*. To be precise, let $x_i$ be the $i^{th}$ bit of $x$ and $y_i$ be the $i^{th}$ bit of $y$. For any pair of input strings $(x, y)$, we partition it by telling the value of each $x_i$ and $y_i$ (by sending a message of the form "$x_i = 0$" or "$x_i = 1$") to a random player. As before, we say that a protocol $\mathcal{R}$ has *complexity $t$* if it always uses at most $t$ bits in total, *regardless of the input and its partition*. We note that the error probability of a protocol $\mathcal{R}$ is calculated over all possible random choices made by the algorithm

---

[4]We emphasize that we allow $\mathcal{R}$ to incur at most $t$ bits of communication regardless of the input and random choices made by the protocol. We note a standard fact that one can also define the complexity $t$ to be the *expected* number of bits. The two notions are equivalent up to a constant factor.

*and all possible random partitions*; e.g., it is possible that an $(\epsilon_0, \epsilon_1)$-error protocol never answers a correct answer for some input partition. Also note that, while we pick the input partition randomly, the input pair itself is picked *adversarially*. In other words, an $(\epsilon_0, \epsilon_1)$-error protocol must, for any input $(x, y)$, output 0 with probability at least $1 - \epsilon_0$ when $f(x, y) = 0$ and output 1 with probability at least $1 - \epsilon_1$ when $f(x, y) = 1$, where the probability is over all possible random strings given to the protocol and the random partition. For any function $f$, the *random-partition communication complexity* for computing $f$ with $(\epsilon_0, \epsilon_1)$-error, denoted by $R^{rcc-pub}_{\epsilon_0, \epsilon_1}(f)$, is the minimum $t$ such that there is an $(\epsilon_0, \epsilon_1)$-error protocol with complexity $t$.

The problems of our interest are *equality* and *disjointness*. In the rest of Section 3.1, we show that the communication complexity of these problems are essentially the same in both worst-partition and random-partition models. The techniques used to prove these results are different between the two problems, and might be of an independent interest.

### 3.1.1 ST Verification and Random-Partition Communication Complexity of Eq.

The equality function, denoted by eq, is defined as $\mathsf{eq}(x, y) = 1$ if $x = y$ and $\mathsf{eq}(x, y) = 0$ otherwise. Note that this problem can be solved by the fingerprinting technique which makes a small error only when $x \neq y$, i.e. $R^{cc-pub}_{\epsilon, 0}(\mathsf{eq}) = O(\log b)$ (see, e.g. [22]). Interestingly, if we "switch" the error side, the problem becomes hard: $R^{cc-pub}_{0, \epsilon}(\mathsf{eq}) = \Omega(b)$. We show that this phenomenon remains true in the random-partition setting. Lemma 3.1 is proved in Appendix B.

LEMMA 3.1. (RANDOM-PARTITION EQUALITY) *For some $\epsilon > 0$, $R^{rcc-pub}_{0, \epsilon}(\mathsf{eq}) = \Omega(b)$. This lower bound holds even when Alice knows $x$ and Bob knows $y$.*

**Lower bound for ST verification.** We now show a lower bound of $\tilde{\Omega}(n/k^2)$ on $(0, \epsilon)$-error algorithms for ST verification. For the $b$-bit string inputs $x$ and $y$ of the equality problem, we construct the following graph $G(x, y)$: The nodes are $u_0, \ldots, u_b$ and $v_0, \ldots, v_b$. For any $i$, there is an edge between $u_0$ and $u_i$ if and only if $x_i = 1$, and there is an edge between $v_0$ and $v_i$ if and only if $y_i = 0$. Additionally, there is always an edge between $u_j$ and $v_j$, for $0 \leq j \leq b$. Observe that $G(x, y)$ is a spanning tree if and only if $x = y$. Also note that $G(x, y)$ has $n = \Theta(b)$ nodes.

Now assume that there is a $(0, \epsilon)$-error algorithm $\mathcal{R}$ in the $k$-machine model that finishes in $\tilde{o}(n/k^2)$ rounds. Alice and Bob simulate $\mathcal{R}$ as follows. Let $p_1, \ldots, p_k$ be machines in the $k$-machine model, and assume that $k$ is even. First, Alice and Bob generate a random partition of nodes in $G(x, y)$ using the random partition of input $(x, y)$ and shared randomness. Using shared randomness, they decide which machine the nodes $u_0$ and $v_0$ should belong to. Without loss of generality, we can assume that $u_0$ belongs to $p_1$. Moreover, with probability $1 - 1/k$, $v_0$ is not in $p_1$, and we can assume that $v_0$ is in $p_2$ without loss of generality. (If $u_0$ and $v_0$ are in the same machine then Alice and Bob stop the simulation and output 0 (i.e. $x \neq y$).) Alice will simulate machines in $P_A = \{p_1, p_3, p_5, \ldots, p_{k-1}\}$ and Bob will simulate machines in $P_B = \{p_2, p_4, \ldots, p_k\}$. This means that Alice (Bob respectively) can put and get any information from $P_A$ ($P_B$ respectively) with no cost. At this point, Alice assigns node $u_0$ on $p_1$; i.e., she tells $p_1$ whether $u_i$ has an edge to $u_0$ or not, for all $i$ (this can be done since she knows $x$). Similarly, Bob assigns node $v_0$ on $p_2$. (Note that to do this we need Alice to know $x$ and Bob to know $y$ in addition to the random partition of $x$ and $y$. We have the lower bound of this as claimed in Lemma 3.1.) Next, they randomly put every node in a random machine. For any $i$, if Alice gets $x_i$, then she assigns $u_i$ in a random machine in $P_A$, i.e., she tells such a random machine whether $u_i$ has an edge to $u_0$ or not. Otherwise, Bob puts $u_i$ in a random machine in $P_B$ in the same way. Since each $x_i$ and $y_i$ belongs to Alice with probability $1/2$, it can be seen that each $u_i$ and $v_i$ will be assigned to a random machine. Similarly, node $v_i$ is assigned to a random machine depending on who gets $y_i$. Note that both Alice and Bob know which machine each node is assigned to since they use shared randomness.

Now Alice and Bob simulate $\mathcal{R}$ where Alice simulates $\mathcal{R}$ on machines in $P_A$ and Bob simulates $\mathcal{R}$ on machines in $P_B$. To keep this simulation going, they have to send messages to each other every time there is a communication between machines in $P_A$ and $P_B$. This means that they have to communicate $\tilde{O}(k^2)$ bits in each round of $\mathcal{R}$. Since $\mathcal{R}$ finishes in $\tilde{o}(n/k^2)$ rounds, Alice and Bob have to communicate $\tilde{o}(n) = \tilde{o}(b)$ bits. Once they know whether $G(x, y)$ is an ST or not, they can answer whether $x = y$ or not. Since $\mathcal{R}$ is $(0, \epsilon)$-error, Alice and Bob's error will be $(0, \epsilon + 1/k)$, where the extra $1/k$ term is because they answer 0 when $u_0$ and $v_0$ are in the same machine. For large enough $k$, this error is smaller than the error in Lemma 3.1, contradicting Lemma 3.1. This implies that such an algorithm $\mathcal{R}$ does not exist.

### 3.1.2 Conn and Random-Partition Communication Complexity of Disjointness

The disjointness function, denoted by disj, is defined as $\mathsf{disj}(x, y) = 1$ if there is $i$ such that $x_i = y_i$ and $\mathsf{disj}(x, y) = 0$ otherwise. This problem in the worst-partition model is a fundamental problem in communication complexity, having tons of application (e.g. [9]). Through a series of results (e.g. [2, 4, 7, 18, 36]), it is known that $R^{cc-pub}_{\epsilon,\epsilon}(\text{disj}) = \Omega(b)$. By adapting the previous proof of Razborov [36], we show that this lower bound remains true in the random-partition setting. Lemma 3.2 is proved in Appendix A.

LEMMA 3.2. (RANDOM-PARTITION DISJOINTNESS) *For some $\epsilon > 0$, $R^{rcc-pub}_{\epsilon,\epsilon}(\text{disj}) = \Omega(b)$. This lower bound holds even when Alice knows $x$ and Bob knows $y$.*

**Lower bound for Conn verification.** We now show a lower bound of $\tilde{\Omega}(n/k^2)$ on $(\epsilon, \epsilon)$-error algorithms for Conn verification, for a small enough constant $\epsilon > 0$. For the $b$-bit string inputs $x$ and $y$ of the disjointness problem, we construct the following graph $G(x, y)$: The nodes are $u_0, \ldots, u_b$ and $v_0, \ldots, v_b$. For any $i$, there is an edge between $u_0$ and $u_i$ if and only if $x_i = 0$, and there is an edge between $v_0$ and $v_i$ if and only if $y_i = 0$. Additionally, there is always an edge between $u_j$ and $v_j$, for $0 \leq j \leq b$. Observe that $G(x, y)$ is connected if and only if $x$ and $y$ are disjoint. Also note that $G(x, y)$ has $n = \Theta(b)$ nodes. Assume that there is an $(\epsilon, \epsilon)$-error algorithm $\mathcal{R}$ in the $k$-machine model that finishes in $\tilde{o}(n/k^2)$ rounds. Alice and Bob simulate $\mathcal{R}$ as follows. Let $p_1, \ldots, p_k$ be machines in the $k$-machine model, and assume that $k$ is even. First, Alice and Bob generate a random partition of nodes in $G(x, y)$ using the random partition of input $(x, y)$ and shared randomness. Using shared randomness, they decide which machine the nodes $u_0$ and $v_0$ should belong to. Without loss of generality, we can assume that $u_0$ belongs to $p_1$. Moreover, with probability $1 - 1/k$, $v_0$ is not in $p_1$, and we can assume that $v_0$ is in $p_2$ without loss of generality. (If $u_0$ and $v_0$ are in the same machine then Alice and Bob stop the simulation and output 0.) Alice will simulate machines in $P_A = \{p_1, p_3, p_5 \ldots, p_{k-1}\}$ and Bob will simulate machines in $P_B = \{p_2, p_4, \ldots, p_k\}$. This means that Alice (Bob respectively) can put and get any information from $P_A$ ($P_B$ respectively) with no cost. At this point, Alice assigns node $u_0$ on $p_1$; i.e., she tells $p_1$ whether $u_i$ has an edge to $u_0$ or not, for all $i$ (this can be done since she knows $x$). Similarly, Bob assigns node $v_0$ on $p_2$. Next, they randomly put every node in a random machine. For any $i$, if Alice gets $x_i$, then she assigns $u_i$ in a random machine in $P_A$, i.e., she tells such a random machine whether $u_i$ has an edge to $u_0$ or not. Otherwise, Bob puts $u_i$ in a random machine in $P_B$ in the same way. Since each $x_i$ and $y_i$ belongs to Alice with probability $1/2$, it can be seen that each $u_i$ and $v_i$ will be assigned to a random machine. Similarly, node $v_i$ is assigned to a random machine depending on who gets $y_i$. Note that both Alice and Bob knows which machine each node is assigned to since they use shared randomness.

Now Alice and Bob simulate $\mathcal{R}$ where Alice simulates $\mathcal{R}$ on machines in $P_A$ and Bob simulates $\mathcal{R}$ on machines in $P_B$. To keep this simulation going, they have to send messages between each other every time there is a communication between machines in $P_A$ and $P_B$. This means that they have to communicate $\tilde{O}(k^2)$ bits in each round of $\mathcal{R}$. Since $\mathcal{R}$ finishes in $\tilde{o}(n/k^2)$ rounds, Alice and Bob have to communicate $\tilde{o}(n) = \tilde{o}(b)$ bits. Once they know whether $G(x, y)$ is connected or not, they can answer whether $x$ and $y$ are disjoint or not. Since $\mathcal{R}$ is $(\epsilon, \epsilon)$-error, Alice and Bob's error will be $(\epsilon, \epsilon + 1/k)$, where the extra $1/k$ term is because they answer 0 when $u_0$ and $v_0$ are in the same machine. For large enough $k$, this error is smaller than the error in Lemma 3.2, contradicting Lemma 3.2. This implies that such an algorithm $\mathcal{R}$ does not exist.

## 4 Algorithms and Techniques

**4.1 A Mapping Lemma** We first prove a "mapping" lemma for the random vertex partition model, which we will use in our Conversion theorem (cf. Theorem 4.1). Consider an input graph $G = (V, E)$ that is partitioned (according to the random vertex partition model) among the $k$ machines in $N = \{p_1, \ldots, p_k\}$ of the network. Let $|V| = n$ and $|E| = m$. We will assume (without loss of generality) throughout that $n \geq k$. We say that a vertex $v$ of $G$ is mapped to a machine $h$ of $N$ if $v$ is assigned to $N$, i.e., $h$ is the home machine of $v$ (cf. Section 1.1). We say that an edge $e = (u, v)$ of $G$ is *mapped* to a link $(p_i, p_j)$ of the $k$-machine network, if $u$ is mapped to $p_i$ and $v$ is mapped to $v_j$ or vice versa. The following Mapping Lemma gives a concentration bound on the number of edges mapped to any link of the network.

LEMMA 4.1. (MAPPING LEMMA) *Let an $n$-node, $m$-edge graph $G$ be partitioned among the $k$ machines in $N = \{p_1, \ldots, p_k\}$, according to the random vertex partition model (assume $n \geq k$). Then with probability at least $1 - 1/n^\alpha$, where $\alpha > 1$ is an arbitrary fixed constant, the following bounds hold:*
**(1)** *The number of vertices of $G$ mapped to any machine is $\tilde{O}(n/k)$.*
**(2)** *The number of edges of $G$ mapped to any link of the network is $\tilde{O}(m/k^2 + \Delta/k)$, where $\Delta$ is the maximum node degree of $G$.*

*Proof.* **(1)** This follows easily from a direct Chernoff bound application. Since each vertex of $G$ is mapped independently and uniformly to the set of $k$ machines, the expected number of vertices mapped to a machine is $n/k$. The concentration follows from a standard Chernoff bound [14].

**(2)** We first note that we cannot directly apply a Chernoff bound to show concentration on the number of edges mapped, as these are not independently distributed. We show the bound in two steps:
1. (a) We first show a concentration bound on the total degree of the vertices assigned to any machine;
2. (b) then we bound the number of edges assigned to any link.

To show (a) we use *Bernstein's inequality* [14]. Fix a machine $p$. Let random variable $X_i^p$ be defined as follows: $X_i^p = d(v_i)$ ($d(v_i)$ is the degree of $v_i$) if vertex $v_i$ is assigned to machine $p$, otherwise $X_i^p = 0$. Let $X^p = \sum_{i=1}^n X_i^p$ denote the total degree of the vertices assigned to machine $p$; in other words, it is the total number of edges that have at least one endvertex assigned to machine $p$. We have $E[X_i^p] = d(v_i)/k$ and $E[X^p] = \sum_{i=1}^n E[X_i^p] = \sum_{i=1}^n d(v_i)/k = 2m/k$. Furthermore, $Var(X_i^p) = E[(X_i^p)^2] - E[X_i^p]^2 = \frac{1}{k}(d(v_i))^2 - (\frac{d(v_i)}{k})^2 = \frac{(d(v_i))^2}{k}(1 - 1/k)$ and hence $Var(X^p) = \sum_{i=1}^n Var(X_i^p) = \frac{1}{k}(1-\frac{1}{k})\sum_{i=1}^n (d(v_i))^2 \leq \frac{1}{k}(1-\frac{1}{k})\sum_{i=1}^n \Delta d(v_i) = \frac{1}{k}(1-\frac{1}{k})\Delta m$.

Using Bernstein's inequality, we have (for some $t > 0$):

$$\Pr(X^p > E[X^p] + t) \leq e^{-\frac{t^2}{2Var(X^p)+(2/3)bt}}$$

where $b = \max_{1 \leq i \leq n}|X_i^p - E[X_i^p]|$. Now, $|X_i^p - E[X_i^p]| \leq d(v_i)(1-1/k) \leq \Delta(1-1/k) = b$.

Let $\gamma > 0$ and let $A$ be the event that $X^p > 2m/k + \gamma(2m/k + \Delta)$. Hence, for any $\gamma > 0$ and letting $t = \gamma(2m/k + \Delta)$, we have:

$$\Pr(A) \leq e^{-\frac{t^2}{2(1/k)(1-1/k)\Delta m+(2/3)\Delta(1-1/k)t}} \leq e^{-\frac{t^2}{\Theta(m\Delta/k+\Delta t)}}$$
$$\leq e^{-\frac{\gamma^2(2m/k+\Delta)^2}{\Theta(m\Delta/k+\Delta^2)}} \leq e^{-\gamma^2\Theta(\frac{m}{\Delta k}+\frac{\Delta k}{m}+1)} = O(1/n^{3\alpha}),$$

if $\gamma = \Theta(\alpha\sqrt{\log n})$.

The above tail bound applies to a single machine $p$; applying a union bound over all the $k$ machines, we have $X^p = \tilde{O}(m/k + \Delta)$ whp for every machine $p \in N$.

We now show (b) which will complete the proof of Part 2. Fix a link $\ell = (p,q)$ of the $k$-machine network. Let $X^p$ (resp. $X^q$) be defined as before, i.e., the total number of edges of $G$ with at least one endvertex assigned to $p$ (resp. $q$). Note that each such edge mapped to $p$ has probability of $1/(k-1)$ of being mapped to link $\ell$ (independently of other edges mapped to $p$). A similar statement holds for edges mapped to $q$ as well. Let r.v. $Z_\ell^p$ (resp. $Z_\ell^q$) denote the number of edges, among those mapped to $p$ (resp. $q$), that are mapped to link $\ell$; the total number of edges mapped to $\ell$ is bounded by $Z_\ell^p + Z_\ell^q$. We have, $E[Z_\ell^p|X^p] \leq \lceil \frac{X^p}{k-1} \rceil$ and similarly $E[Z_\ell^q|X^q] \leq \lceil \frac{X^q}{k-1} \rceil$. We next show that $Z_\ell^p$ and $Z_\ell^q$ are both bounded by $\tilde{O}(m/k^2 + \Delta/k)$ whp which will complete the proof. We focus on $Z_\ell^p$; (the case of $Z_\ell^q$ is similar). By proof of (a), $X^p < \Phi$ with probability at least $1 - 1/n^{2\alpha}$, where $\Phi = c\log n(m/k + \Delta)$ for some sufficiently large constant $c > 0$, depending on constant $\alpha$. Observe that an edge that has one endpoint mapped to $p$ is mapped to link $\ell$ independently. Hence we can apply a standard Chernoff bound [14]: $\Pr(Z_\ell^p < c'\Phi/k + 1) \leq \Pr(X^p > \Phi) + \Pr(Z_\ell^p < c'\Phi/k + 1)|X^p < \Phi) \leq 1/n^{2\alpha} + 2^{-c'\Phi/k+1} \leq 1/n^{2\alpha} + 2^{-3\log n+\Omega(1)} = O(1/n^{2\alpha})$, for a sufficiently large constant $c' > 0$.

Thus, with probability $1 - 1/n^{2\alpha}$, the number of edges mapped to link $\ell$ is $c'\Phi/k = \tilde{O}(m/k^2 + \Delta/k)$. Applying a union bound over all $k(k-1)/2$ links, yields the result.

**4.2 The Conversion Theorem** We now present a general conversion theorem that enables us to leverage results from the standard message-passing model [33]. Our conversion theorem allows us to use distributed algorithms that leverage direct communication between nodes, even when such an edge is not part of the input graph. More specifically, we can translate any distributed algorithm that works in the following clique model to the $k$-machine model.

**The Clique Model.** Consider a complete $n$-node network $C$ and a spanning subgraph $G$ of $C$ determined by a set of (possibly weighted) edges $E(G)$. The nodes of $C$ execute a distributed algorithm and each node $u$ is aware of the edges that are incident to $u$ in $G$. Each node can send a message of at most $W \geq 1$ bits over each incident link per round. For a graph problem $P$, we are interested in distributed algorithms that run on the network $C$ and, given input graph $G$, compute a feasible solution of $P$. In addition to *time complexity* (the number of rounds in the worst case), we are interested in the *message complexity* of an algorithm in this model which is the number of messages (in the worst case) sent over all links. Additionally, we are also interested in *communication degree complexity* which is the maximum number of messages sent or received by any node in any round; i.e., it is the minimum integer $M'$ such that every node sends a message to at most $M'$ other nodes in each round. Note that we can simulate any "classic" distributed algorithm running on a network $G$ of an arbitrary topology that uses messages of $O(\log n)$ size in the clique model by simply restricting the communication to edges in $E(G) \subset E(C)$ and by splitting messages into packets of size $W$. In this case, the time and message complexities remain the same (up to log-factors) while the communication degree complexity can be bounded by the maximum degree of $G$. We say that an algorithm is a *broadcast* algorithm if, in every round and for every node $u$, it holds that $u$ broadcasts the same message to other nodes (or remains silent). We define the *broadcast complexity* of an algorithm as the number of times nodes broadcast messages.

THEOREM 4.1. (CONVERSION THEOREM) *Suppose that there is an $\varepsilon$-error algorithm $A_C$ that solves problem $P$ in time $T_C(n) \in \tilde{O}(n)$ in the clique model, for any $n$-node input graph. Then there exists an $\varepsilon$-error algorithm $A$ that solves $P$ in the $k$-machine model with bandwidth $W$ satisfying the following time complexity bounds with high probability:*

**(a)** *If $A_C$ uses point-to-point communication with message complexity $M$ and communication degree complexity $\Delta'$, then $A$ runs in $\tilde{O}\left(\frac{M}{k^2 W} + T_C(n)\lceil\frac{\Delta'}{kW}\rceil\right)$ time.*

**(b)** *If $A_C$ is a broadcast algorithm with broadcast complexity $B$, then $A$ takes $\tilde{O}(\frac{B}{kW} + T_C(n))$ time.*

*Proof.* Consider any $n$-node input graph $G$ with $m$ edges and suppose that nodes in $G$ are assigned to the $k$ machines of the network $N$ according to the vertex partitioning process (cf. Section 1.1).

We now describe how to obtain algorithm $A$ for the $k$-machine model from the clique model algorithm $A_C$: Each machine locally simulates the execution of $A_C$ at each hosted vertex. First of all, we only need to consider inter-machine communication, since local computation at each machine happens instantaneously at zero cost. If algorithm $A_C$ requires a message to be sent from a node $u_1 \in C$ hosted at machine $p_1$ to some node $u_2 \in C$ hosted at $p_2$, then $p_1$ sends this message directly to $p_2$ via the links of the network $N$. (Recall that a machine $p_1$ knows the hosting machines of all endpoints of all edges (in $G$) that are incident to a node hosted at $p_1$.) Moreover, $p_1$ adds a header containing the IDs of $u_1$ and $u_2$ to ensure that $p_2$ can correctly deliver the message to the simulation of $A_C$ at $u_2$. Each message is split into packets of size $W$, which means that sending all packages that correspond to such a message requires $\lceil O(\log n)/W\rceil$ rounds. In the worst case (i.e. $W=1$), this requires additional $O(\log n)$ rounds, which does not change our complexity bounds. Thus, for the remainder of the proof, we assume that $W$ is large enough such that any message generated by $A_C$ can be sent in 1 round in the $k$-machine model.

*Proof of (a):* We will bound the number of messages sent in each round through each link using Lemma 4.1(2).

Let $G_i$ be the graph whose node set is the same as the input graph (as well as the clique model), and there is an edge between nodes $u$ and $v$ if and only if a message is sent between $u$ and $v$ in round $i$ of the algorithm; in other words, $G_i$ captures the communications happening in round $i$. From Lemma 4.1(2), we know that (w.h.p.) each communication link of $N$ is mapped to at most $\tilde{O}(|E(G_i)|/k^2 + \Delta_i/k)$ edges of $G_i$, where $\Delta_i$ is the maximum degree of $G_i$. This means that each machine needs to send at most $\tilde{O}(|E(G_i)|/k^2 + \Delta_i/k)$ messages over a specific communication link with high probability. In other words, the $i^{th}$ round of $A_C$ can be simulated in $\tilde{O}(|E(G_i)|/k^2 W + \Delta_i/kW)$ rounds, and, by taking a union bound, the same is true for all rounds in $[1, T_C(n)]$. By summing up over all rounds of $A_C$, we can conclude that the number rounds needed to simulate $A_C$ is

$$\tilde{O}\left(\sum_{i=1}^{T_C(n)}\left(\frac{|E(G_i)|}{k^2 W} + \frac{\Delta_i}{kW}\right)\right) = \tilde{O}\left(\frac{M}{k^2 W} + T_C(n)\left\lceil\frac{\Delta'}{kW}\right\rceil\right)$$

where the equality is because of the following facts: (1) $\sum_{i=1}^{T_C(n)}|E(G_i)| = O(M)$ since $|E(G_i)|$ is at most two times the number of messages sent by all nodes in the $i^{th}$ round, and (2) $\Delta_i \leq \Delta'$. This proves (a).

*Proof of (b):* We first slightly modify the previous simulation to simulate broadcast algorithms: Note that if $A_C$ is a broadcast algorithm, then for the $i^{th}$ round $(i \geq 1)$ of algorithm $A_C$, if a node $u$ belonging to machine $p_1$ sends messages to nodes $v_1, \ldots, v_j$ $(j \geq 1)$ belonging to machine $p_2$, we know that $u$ sends *the same message* to $v_1, \ldots, v_j$. Thus, when we simulate this round $A_C$, we will let machine $p_1$ send only one message to $p_2$, instead of $j$ messages. Then, machine $p_2$ will pretend that this message is sent from $u_1$ to *all nodes* belonging to $p_2$ that have an edge to node $u$. (We cannot specify the destination nodes $v_1, \ldots, v_j$ in this message as this might increase the length of the message significantly.)

We now analyze this new simulation. We show that this simulation finishes in $\tilde{O}(\frac{B}{k} + T_C(n))$ rounds. Let $B_i$ be the number of nodes that perform a broadcast in round $i$ of the run of $A_C$ in the clique model, and note that $B = \sum_{i=1}^{T_C(n)} B_i$. According to Lemma 4.1(a), the number of nodes contributing to $B_i$ broadcasts that are assigned to a single machine is $\tilde{O}(\lceil B_i/k\rceil)$ w.h.p.; in other words, w.h.p., each machine contains $\ell_i = \tilde{O}(\lceil B_i/k\rceil)$ of the $B_i$ nodes. Thus, for every $i$, we instruct algorithm $A$ to simulate these $B_i$ broadcasts in the $k$-machine model in $\lceil\ell_i/W\rceil$ rounds. Since $A_C$ takes at most $T_C(n)$ rounds, we can take a union bound, and it follows that algorithm $A$ takes $\tilde{O}(\frac{B}{kW} + T_C(n))$ rounds in the $k$-machine model.

It is easy to see that a simulation similar to the one employed in the proof of Theorem 4.1 provides the same complexity bounds, if we limit the total communication of each machine (i.e. bits sent/received) to at most $kW$ bits per round, instead of restricting the bandwidth of individual inter-machine links to $W$ bits. To see why this is true, observe that throughout the above simulation, each machine is required to send/receive at most $kW$ bits in total per simulated round with high probability.

**4.3 Algorithms** We now consider various important graph problems in the $k$-machine model. For the sake of readability, we assume a bandwidth of $\Theta(\log n)$ bits, i.e., parameter $W = \Theta(\log n)$. Observe that the simple solution of aggregating the entire information about the input graph $G$ at a single machine takes $O(m/k)$ rounds; thus we are only interested in algorithms that beat this trivial upper bound. Our results are summarized in Table 1.

**Breadth-First Search Tree (BFS).** To get an intuition for the different bounds obtained by applying either Theorem 4.1(a) or Theorem 4.1(b) to an algorithm in the clique model, consider the problem of computing a breadth-first search (BFS) tree rooted at a fixed source node. If we use Theorem 4.1(a) we get a bound of $\tilde{O}(m/k^2 + D\lceil\Delta/k\rceil)$ rounds. In contrast, recalling that each node performs $O(1)$ broadcasts, Theorem 4.1(b) yields $\mathcal{T}_{1/n}^k(\mathsf{BFS}) \in \tilde{O}(n/k + D)$. We will leverage these bounds when considering graph connectivity and spanning tree verification below.

**Minimum Spanning Tree (MST), Spanning Tree Verification (ST) and Graph Connectivity (Conn).** An efficient algorithm for computing the MST of an input graph was given by [16], which proceeds by merging "MST-fragments" in parallel; initially each vertex forms a fragment by itself. In each of the $O(\log n)$ phases, each fragment computes the minimum outgoing edge (pointing to another fragment) and tries to merge with the respective fragment. Since any MST has $n-1$ edges, at most $n-1$ edges need to be added in total. This yields a total broadcast complexity of $\tilde{O}(n)$ and thus Theorem 4.1(b) readily implies the bound of $\tilde{O}(n/k)$. We can use an MST algorithm for verifying *graph connectivity* which in turn can be used for ST. We assign weight 1 to all edges of the input graph $G$ and then add an edge with infinite weight between any pair of nodes $u$, $v$ where $(u,v) \notin E(G)$, yielding a modified graph $G'$. Clearly, $G$ is disconnected iff an MST of $G'$ contains an edge with infinite weight. This yields the first part of the upper bound for graph connectivity stated in Table 1. We now describe how to verify whether an edge set $S$ is an ST, by employing a given algorithm $A$ for Conn. Note that, for ST *verification*, each machine $p$ initially knows the assumed status of the edges incident to its nodes wrt. being part of the ST, and eventually $p$ has to output either YES or NO. First, we run $A$ on the graph induced by $S$ and then we compute the size of $S$ as follows: Each machine locally adds 1 to its count for each edge $(u,v) \in S$, if $p$ is the home machine for vertices $u$, $v$. Otherwise, if one of $u$ or $v$ reside on a different machine, then $p$ adds $1/2$. Then, all machines exchange their counts via broadcast, which takes 1 round (since each count is at most $n$ and $W \in \Theta(\log n)$) and determine the final count by summing up over all received counts including their own. Each machine outputs YES iff (1) the output of the Conn algorithm $A$ returned YES and (2) the final count is $n - 1$. Thus we get the same bounds for ST verification as for graph connectivity.

Recalling that we can compute a BFS in $\tilde{O}(m/k^2 + D\lceil\Delta/k\rceil)$ rounds, it is straightforward to see that the same bound holds for Conn (and thus also ST verification): First, we run a leader election algorithm among the $k$ machines. This can be done in $O(1)$ rounds (and $\tilde{O}(\sqrt{k})$ messages) by using the algorithm of [23] (whp). The designated leader machine then chooses an arbitrary node $s$ as the source node and executes a BFS algorithm. Once this algorithm has terminated, each machine locally computes the number of its vertices that are part of the BFS and then computes the total number of vertices in the BFS by exchanging its count (similarly to the ST verification above). The input graph is connected iff the BFS contains all vertices.

**PageRank.** The PageRank problem is to compute the PageRank distribution of a given graph (may be directed or undirected). A distributed page rank algorithm was presented in [10], based on the distributed random walk algorithm: Initially, each node generates $\Theta(\log n)$ random walk tokens. A node forwards each token with probability $1-\delta$ and terminates the token with probability $\delta$ (called the *reset* probability). Clearly, every token will take at most $O(\log n/\delta)$ steps with high probability before being terminated. From Lemma 2.2 of [11] we know that these steps can be implemented in $O(\log^2 n)$ rounds in the clique model, and since this requires $O(n \log^2 n/\delta)$ messages to be sent in total, Theorem 4.1(a) yields that, for any $\delta > 0$, there is a randomized algorithm for computing PageRank in the $k$-machine model such that $\mathcal{T}_{1/n}^k(\mathsf{PageRank}) \in \tilde{O}(\frac{n}{\delta k})$.

**Computing a $(2\delta-1)$-Spanner.** The algorithm of [5] computes a $(2\delta-1)$-spanner, for some $(\delta \in O(\log n))$, of the input graph $G$ in $\delta^2$rounds (using messages of $O(\log n)$ size) that has an expected number of $\delta m^{1+1/\delta}$ edges (cf. Theorem 5.1 in [5]). That is, each node needs to broadcast $\delta^2$ times, for a total broadcast complexity of $n\delta^2$. Applying Theorem 4.1(b) yields a bound of $\tilde{O}(n/k)$ rounds in the $k$ machine model.

**Single-Source Shortest Paths (SSSP, SPT), and All-Pairs Shortest Paths(APSP).** We show that, in the $k$-machine model, SSSP and APSP can be $(1+\epsilon)$-approximated in $\tilde{O}(n/\sqrt{k})$ time, and $(2+\epsilon)$-approximated in $\tilde{O}(n\sqrt{n}/k)$ time, respectively.

Recall that, for SSSP, we need to compute the distance between each node and a designated source node. Nanongkai [31] presented a $\tilde{O}(\sqrt{n}D^{1/4})$-time algorithm for SSSP, which implies a $\tilde{O}(\sqrt{n})$-time algorithm in the clique model. We show that the ideas in [31], along with Theorem 4.1(b), leads to a $\tilde{O}(n/\sqrt{k})$-time $(1+\epsilon)$-approximation algorithm in the $k$-machine model. We sketch the algorithm in [31] here[5]. First, every node broadcasts $\rho$ edges incident to it of minimum weight (breaking tie arbitrarily), for some parameter $\rho$ which will be fixed later. Using this information, every node internally compute $\tilde{O}(1)$ integral weight functions (without communication). For each of these weight functions, we compute a BFS tree of depth $n/\rho$ from the source node, treating an edge of weight $w$ as a path of length $w$. Using two techniques called *light-weight SSSP* and *shortest-path diameter reduction*, the algorithm of [31] gives a $(1+\epsilon)$-approximation solution. Observe that this algorithm uses broadcast communication. Its time complexity is clearly $T_C(n) = \tilde{O}(\rho + n/\rho)$. (Thus, by setting $\rho = \sqrt{n}$, we have the running time of $\tilde{O}(\sqrt{n})$ in the clique model.) Its broadcast complexity is $B = \tilde{O}(n\rho)$ since every node has to broadcast $\rho$ edges in the first step and the BFS tree algorithm has $O(n)$ broadcast complexity. Its message complexity is $M = \tilde{O}(n^2\rho)$ (the BFS tree algorithm has $O(n^2)$ message complexity since a message will be sent through each edge once). By Theorem 4.1(b), we have that in the $k$-machine model, the time we need to solve SSSP is $\tilde{O}(\frac{n\rho}{k} + \rho + n/\rho)$. Using $\rho = \sqrt{k}$, we have the running time of $\tilde{O}(\sqrt{k} + n/\sqrt{k}) = \tilde{O}(n/\sqrt{k})$ where the equality is because $k \leq n$.

In [31], a similar idea was also used to obtain a $(2+\epsilon)$-approximation $\sqrt{n}$-time algorithm for APSP on the clique model. This algorithm is almost identical to the above algorithm except that it creates BFS trees of depth $n/\rho$ from $n/\rho$ centers instead of just the source. By this modification, it can be shown that the running time remains the same, i.e. $T_C(n) = \tilde{O}(\rho + n/\rho)$. (Thus, by setting $\rho = \sqrt{n}$, we have the running time of $\tilde{O}(\sqrt{n})$ in the clique model.) The broadcast complexity

---
[5]The algorithm is in fact a simplification of the algorithm in [31] since we only have to deal with the clique model

becomes $B = \tilde{O}(n\rho + n^2/\rho)$ since each BFS tree algorithm has $O(n)$ broadcast complexity. Its message complexity becomes $M = \tilde{O}(n^2\rho + n^3/\rho)$ since each BFS tree algorithm has $O(n^2)$ message complexity. By Theorem 4.1(b), we have that in the $k$-machine model, the time we need to solve APSP is $\tilde{O}(\frac{n\rho+n^2/\rho}{k} + \rho + n/\rho)$. Using $\rho = \sqrt{n}$, we have the running time of $\tilde{O}(\frac{n\sqrt{n}}{k})$.

Since the algorithm of [31] also constructs a shortest path tree while computing the shortest path distances, we get analogous bounds for the SPT problem, which requires each machine to know which of its edges are part of the shortest path tree to the designated source.

We can leverage the technique of computing a $(2\delta-1)$-spanner in $\tilde{O}(n/k)$ rounds that has an expected number of $\tilde{O}(m^{1+1/\delta})$ edges. We can simply collect all edges at one designated machine $p$, which takes time $\tilde{O}(m^{1+1/\delta}/k)$ and then locally compute a $(2\delta-1)$-approximation for the shortest path problems at machine $p$. In particular, for $\delta = \Theta(\log n)$, we have a spanner of (expected) $O(n)$ edges and thus we get a $O(\log n)$-approximation for the shortest path problems in expected $\tilde{O}(n/k)$ rounds.

For computing the exact APSP (resp. SSSP) problems, we can use a distributed algorithm by Bellman-Ford ([33, 28]). This algorithm takes $S$ rounds, where $S$ is the shortest path diameter, and thus the broadcast complexity is $nS$. By virtue of Theorem 4.1(b), we get a round complexity of $\tilde{O}(nS/k + S)$ in the $k$-machine model.

**Densest Subgraph.** We show that Theorem 4.1 implies that the densest subgraph problem can be approximated in $\tilde{O}(\min(\frac{m}{k^2}, \frac{n}{k}))$ time in the $k$-machine model. In [38], Das Sarma et al. presented a $(2+\epsilon)$-approximation algorithm for the densest subgraph problem. The idea is very simple: In every round, they compute the average degree of the network and delete nodes of degree less than $(1+\epsilon)$ times of the average degree. This process generates several subgraphs of the original network, and the algorithm output the densest subgraph among the generated subgraphs. It was shown in [38] that this algorithm produces a $(2+\epsilon)$-approximate solution. They also proved that this algorithm stops after $\log_{1+\epsilon} n$ rounds, implying a time complexity of $T_C(n) = O(\log n)$ in the clique model. The message complexity is $M = \tilde{O}(m)$ since every node has to announce to its neighbors in the input graph that it is deleted at some point. For the same reason, the broadcast complexity is $B = \tilde{O}(n)$. Note that this algorithm is broadcast. So, By Theorem 4.1, we have that in the $k$-machine model, the time we need to solve this problem is $\tilde{O}(\frac{n}{k})$ as desired.

**Cut Sparsifier, Min Cut, Sparsest Cut, etc..** An

$\epsilon$-cut-sparsification of a graph $G$ is a graph $G'$ on the same set of nodes such that every cut in $G'$ is within $(1 \pm \epsilon)$ of the corresponding cut in $G$. We show that we can use the technique called *refinement sampling* of Goel et al. [17], to compute an $\epsilon$-sparsification of $\tilde{O}(n)$ edges in $\tilde{O}(n/k)$ time in the $k$-machine model. By aggregating this sparsification to a single machine, we can approximately solve cut-related problems such as a $(1 \pm \epsilon)$-approximate solution to the minimum cut and sparsest cut problems. The main component of the algorithm of Goel et al. is repetitively sparsify the graph by keeping each edge with probability $2^{-\ell}$ for some $\ell$, and compute the connected components after every time we sparsify the graph. By doing this process for $\tilde{O}(1)$ times, we can compute a probability $z(e)$ on each edge $e$. Goel et al. showed that we can use this probability $z(e)$ to (locally) sample edges and assign weights to them to obtain an $\epsilon$-sparsification. It is thus enough to be able to compute the connected components quickly in the $k$-machine model. This can be done by simply invoke the MST algorithm, which takes $\tilde{O}(n/k)$ rounds. We have to runs this algorithm for $\tilde{O}(1)$ times, so the total running time is $\tilde{O}(n/k)$.

**Covering Problems on Graphs.** We now describe how to solve covering problems like maximal independent set (MIS) in our model. We first consider MIS and related covering problems on simple graphs, and then describe how to obtain an MIS on an input hypergraph. A well known distributed algorithm for computing a maximal independent set (MIS) is due to [27]: The algorithm proceeds in phases and in each phase, every *active* node $v$—initially every node—marks itself with probability $1/2d_v$ where $d_v$ is the degree of $v$. If $v$ turns out to be the only marked node in its neighborhood, $v$ enters the MIS, notifies all of its neighbors who no longer participate (i.e. become *inactive*) in future phases and terminates. When 2 neighboring nodes both mark themselves in the same phase, the lower degree node unmarks itself. Nodes that were not deactivated proceed to the next phase and so forth. It was shown in [27] that this algorithm terminates in $O(\log n)$ rounds with high probability. Since each node sends the same messages to all neighbors, we can analyze the communication in terms of broadcasts, yielding a broadcast complexity of $O(n \log n)$ (whp). Applying Theorem 4.1(b) yields a round complexity of $\tilde{O}(n/k)$. Alternatively, for bounded degree graphs, applying Theorem 4.1(a) gives us a running time of $\tilde{O}(m/k^2 + \Delta/k)$, which is faster when $\Delta \ll k$. Considering the locality preserving reductions (cf. [21]) between MIS, maximal matching (MM), minimal dominating set (MDS), and computing a 2-approximation of the minimum vertex cover (MVC), we get that $\mathcal{T}_{1/n}^k(\mathsf{MIS})$, $\mathcal{T}_{1/n}^k(\mathsf{MM})$, $\mathcal{T}_{1/n}^k(\mathsf{MVC})$, $\mathcal{T}_{1/n}^k(\mathsf{MDS})$ are $\tilde{O}(\min(n/k, m/k^2 + \Delta/k))$.

We now describe how to obtain an $\tilde{O}(n/k+k)$ time algorithm for all graph covering problems directly in the $k$-machine model, without resorting to Theorem 4.1; note that this translates to $\tilde{O}(n/k)$ when $k \leq \tilde{O}(\sqrt{n})$. In particular, this yields an $\tilde{O}(n/k + k)$ algorithm for the problem of finding a maximal independent set in a hypergraph (HMIS), which has been studied extensively in the PRAM model of computation (cf. [20, 6]). Note that, for the hypergraph setting, the input graph $G$ is a hypergraph and if some node $u$ has home machine $p_1$, then $p_1$ knows all hyperedges (and the corresponding machines) that contain $u$. To the best of our knowledge, there is no efficient distributed algorithm known for HMIS. First assume that there is an ordering of the machines with ids $1, \ldots, k$. Such an ordering can be obtained by running the $O(1)$-time leader election algorithm of [23]. The elected leader (machine) then arbitrarily assigns unique ids to all the other machines. Then, we sequentially process the nodes at each machine and proceed in $k$ phases. In the first phase, machine 1 locally determine the membership status of being in the HMIS, for all of its nodes. Next, machine 1 computes an arbitrary enumeration of its nodes and sends the status (either 0 or 1) and node id of the first $k$ nodes over its $k$ links (i.e. a single status is sent over exactly 1 link). When a machine receives this message from machine 1, it simply broadcasts this message to all machines in the next round. Thus, after 2 rounds, all machines know the status of the first $k$ nodes of machine 1. Then, machine 1 sends the status of the next $k$ nodes and so forth. By Lemma 4.1, there will be $\tilde{O}(n/k)$ nodes with high probability, and therefore every machine will know the status of all the nodes of machine 1 after $\tilde{O}(n/k^2)$ rounds. After machine 1 has completed sending out all statuses, all other machines locally use this information to compute the statuses of their nodes (if possible). For example, if some node $u$ at machine 1 is in the HMIS (i.e. has status 1) and adjacent to some node $v$ at machine 2, then machine 2 sets the status of $v$ to 0. Then, machine 2 locally computes the status of its remaining undetermined nodes such that they are consistent with the previously received statuses, and starts sending this information to all other machines in the same fashion. Repeating the same process for each of the $k$ machines yields a total running time of $\tilde{O}(n/k + k)$.

**Finding Triangles and Subgraphs.** For the subgraph isomorphism problem $\mathsf{SubIso}_d$, we are given 2 input graphs: the usual $n$-node graph $G$ and a $d$-vertex graph $H$, for $d \in O(1)$. We want to answer the question whether $H \subseteq G$. A distributed algorithm for the clique model that runs in $O(n^{(d-2)/d})$ rounds was given by [13]. Since the total number of mes-sages sent per round is $O(n^2)$, Theorem 4.1(a) gives rise to an algorithm for the $k$-machine model that runs in $\tilde{O}(n^{2+(d-2)/d}/k^2 + n^{1+(d-2)/d}/k)$ rounds.

We use Tri to denote the restriction of $\mathsf{SubIso}_3$ to the case where $H$ is a triangle. The following is a simple algorithm for the clique model: Each node locally collects its 2-neighborhood information, checks for triangles and then either outputs YES or NO. This requires each node to send a message for each of its at most $\Delta$ neighbors. The total number of messages sent is $n\Delta^2$, and the algorithm takes $\Delta$ rounds to send all messages. Applying Theorem 4.1(a), we get a distributed algorithm in the $k$-machine model with round complexity of $\tilde{O}(n\Delta^2/k^2 + \Delta\lceil \Delta/k\rceil)$, which is better than the above bound when $\Delta$ is sufficiently small. Thus, we have $\mathcal{T}_{1/n}^k(\mathsf{Tri}) \in \tilde{O}(\min(n\Delta^2/k^2 + \Delta\lceil \Delta/k\rceil, n^{7/3}/k^2 + n^{4/3}/k))$.

## 5  Conclusion

We presented algorithms and lower bounds for distributed computation of several graph problems. Our bounds are (almost) tight for problems such as computing a ST or a MST, while for other problems such as connectivity and shortest paths, there is a non-trivial gap between upper and lower bounds. Understanding these bounds and investigating the best possible can provide insight into understanding the complexity of distributed graph processing.

## References

[1] Gps: A graph processing system, http://infolab.stanford.edu/gps/. 1, 2

[2] L. Babai, P. Frankl, and J. Simon. Complexity classes in communication complexity theory (preliminary version). In *FOCS*, pages 337–347, 1986. 9

[3] Nikhil Bansal, Kang-Won Lee, Viswanath Nagarajan, and Murtaza Zafer. Minimum congestion mapping in a cloud. In *PODC*, pages 267–276, 2011. 2

[4] Ziv Bar-Yossef, T. S. Jayram, Ravi Kumar, and D. Sivakumar. An information statistics approach to data stream and communication complexity. *J. Comput. Syst. Sci.*, 68(4):702–732, 2004. Also in FOCS'02. 9

[5] Surender Baswana and Sandeep Sen. A simple and linear time randomized algorithm for computing sparse spanners in weighted graphs. *Random Struct. Algorithms*, 30(4):532–563, 2007. 12, 13

[6] Paul Beame and Michael Luby. Parallel search for maximal independence given minimal dependence. In *SODA*, pages 212–218, 1990. 14

[7] Mark Braverman, Ankit Garg, Denis Pankratov, and Omri Weinstein. From information to exact communication. In *STOC*, pages 151–160, 2013. 9

[8] Amit Chakrabarti, Graham Cormode, and Andrew McGregor. Robust lower bounds for communication and stream computation. *Electronic Colloquium on Computational Complexity (ECCC)*, 18:62, 2011. 3

[9] Arkadev Chattopadhyay and Toniann Pitassi. The Story of Set Disjointness. *SIGACT News*, 41(3):59–85, 2010. 9

[10] Atish Das Sarma, Anisur Rahaman Molla, Gopal Pandurangan, and Eli Upfal. Fast distributed pagerank computation. In *ICDCN*, pages 11–26, 2013. 12

[11] Atish Das Sarma, Danupon Nanongkai, and Gopal Pandurangan. Fast distributed random walks. In *PODC*, pages 161–170, 2009. 12

[12] Jeffrey Dean and Sanjay Ghemawat. Mapreduce: Simplified data processing on large clusters. In *OSDI*, pages 137–150, 2004. 1

[13] Danny Dolev, Christoph Lenzen, and Shir Peled. "tri, tri again": Finding triangles and small subgraphs in a distributed setting - (extended abstract). In *DISC*, pages 195–209, 2012. 14

[14] Dubhashi and Panconesi. *Concentration of Measure for Analysis of Randomized Algorithms*. Cambridge University Press, 2012. 10

[15] Michael Elkin, Hartmut Klauck, Danupon Nanongkai, and Gopal Pandurangan. Can quantum communication speed up distributed computation? In *PODC*, 2014. 5

[16] Robert G. Gallager, Pierre A. Humblet, and Philip M. Spira. A distributed algorithm for minimum-weight spanning trees. *ACM Trans. Program. Lang. Syst.*, 5(1):66–77, 1983. 12

[17] Ashish Goel, Michael Kapralov, and Sanjeev Khanna. Graph sparsification via refinement sampling. *CoRR*, abs/1004.4915, 2010. 14

[18] Bala Kalyanasundaram and Georg Schnitger. The Probabilistic Communication Complexity of Set Intersection. *SIAM J. Discrete Math.*, 5(4):545–557, 1992. Also in CCC'87. 9

[19] Howard J. Karloff, Siddharth Suri, and Sergei Vassilvitskii. A model of computation for mapreduce. In *SODA*, pages 938–948, 2010. 4

[20] Pierre Kelsen. On the parallel complexity of computing a maximal independent set in a hypergraph. In *STOC*, pages 339–350, 1992. 14

[21] Fabian Kuhn, Thomas Moscibroda, and Roger Wattenhofer. Local computation: Lower and upper bounds. *CoRR*, abs/1011.5470, 2010. 14

[22] E. Kushilevitz and N. Nisan. *Communication complexity*. Cambridge University Press, New York, NY, USA, 1997. 7, 8, 20

[23] Shay Kutten, Gopal Pandurangan, David Peleg, Peter Robinson, and Amitabh Trehan. Sublinear bounds for randomized leader election. In *ICDCN*, pages 348–362, 2013. 12, 14

[24] Silvio Lattanzi, Benjamin Moseley, Siddharth Suri, and Sergei Vassilvitskii. Filtering: a method for solving graph problems in mapreduce. In *SPAA*, pages 85–94, 2011. 4

[25] J. Lin and C. Dyer. *Data-Intensive Text Processing with MapReduce.* Morgan Claypool Publishers, 2010. 1, 4

[26] Yucheng Low, Joseph Gonzalez, Aapo Kyrola, Danny Bickson, Carlos Guestrin, and Joseph M. Hellerstein. Graphlab: A new framework for parallel machine learning. In *UAI*, pages 340–349, 2010. 1

[27] Michael Luby. A simple parallel algorithm for the maximal independent set problem. *SIAM J. Comput.*, 15(4):1036–1053, 1986. 14

[28] Nancy Lynch. *Distributed Algorithms.* Morgan Kaufmann, 1996. 2, 13

[29] Grzegorz Malewicz, Matthew H. Austern, Aart J. C. Bik, James C. Dehnert, Ilan Horn, Naty Leiser, and Grzegorz Czajkowski. Pregel: a system for large-scale graph processing. In *SIGMOD Conference*, pages 135–146, 2010. 1, 2

[30] M. Mitzenmacher and E. Upfal. *Probability and Computing: Randomized Algorithms and Probabilistic Analysis.* Cambridge University Press, 2005. 20

[31] Danupon Nanongkai. Distributed approximation algorithms for weighted shortest paths. In *STOC*, pages 565–573, 2014. 13

[32] Danupon Nanongkai, Atish Das Sarma, and Gopal Pandurangan. A tight unconditional lower bound on distributed randomwalk computation. In *PODC*, pages 257–266, 2011. 5

[33] David Peleg. *Distributed computing: a locality-sensitive approach.* Society for Industrial and Applied Mathematics, Philadelphia, PA, USA, 2000. 2, 3, 4, 5, 10, 13

[34] The Apache Project. Apache Giraph, http://giraph.apache.org/. 1, 2

[35] Anand Rajaraman and Jeff Ullman. *Mining of Massive Datasets.* Cambridge University Press, October 2011. 2, 4

[36] Alexander A. Razborov. On the Distributional Complexity of Disjointness. *Theor. Comput. Sci.*, 106(2):385–390, 1992. Also in ICALP'90. 3, 9, 16

[37] Atish Das Sarma, Stephan Holzer, Liah Kor, Amos Korman, Danupon Nanongkai, Gopal Pandurangan, David Peleg, and Roger Wattenhofer. Distributed verification and hardness of distributed approximation. *SIAM J. Comput.*, 41(5):1235–1265, 2012. 5

[38] Atish Das Sarma, Ashwin Lall, Danupon Nanongkai, and Amitabh Trehan. Dense subgraphs on dynamic networks. In *Distributed Computing - 26th International Symposium, DISC 2012, Salvador, Brazil, October 16-18, 2012. Proceedings*, pages 151–165, 2012. 13

[39] Isabelle Stanton. Streaming balanced graph partitioning for random graphs. *CoRR*, abs/1212.1121, 2012. 2

[40] Isabelle Stanton and Gabriel Kliot. Streaming graph partitioning for large distributed graphs. In *KDD*, pages 1222–1230, 2012. 2

[41] David P. Woodruff and Qin Zhang. When distributed computation is communication expensive. In *DISC*, pages 16–30, 2013. 3, 4, 5

## A   Proof of Lemma 3.2

Our proof adapts the proof of Razborov [36] from the situation of a fixed partition to a random partition of the inputs. In the following we try to keep our notation close to that of Razborov's. In particular, we use $N$ to replace $b$ as elsewhere in this paper. We first recall the standard definition of deterministic and randomized protocols here.

DEFINITION 1. *The communication cost of a deterministic protocol $P$ computing a function $f : \mathcal{X} \times \mathcal{Y} \to \{0,1\}$ on an input $x, y$ is the number of bits exchanged between Alice and Bob during $P$ on $x, y$. The communication cost of $P$ is the maximum communication cost over all inputs. The deterministic communication complexity of $f$ is the minimum communication cost (over all protocols) computing $f$.*

*A randomized protocol has to be correct with probability $2/3$ (over the random choices used in the protocol) on every input. The randomized communication complexity is then defined as above.*

In the above definition, the inputs $x, y$ are partitioned in a fixed way. We are interested in the communication cost over an average partition of the inputs to Alice and Bob. In this situation the communication cost (for a fixed input $z$) is the average over all partitions of the bits of $z$ to Alice and Bob. More precisely, for every partition $\rho$ of $z$ there is a protocol $P_\rho$ computing $f$, and the communication cost on $z$ is the average over the communication costs of all the $P_\rho$ on $z$. Here the distribution over partitions is the one in which every bit $z_j$ is assigned uniformly random to Alice or Bob (but not both).

DEFINITION 2. *A deterministic average partition protocol $P$ consists of a protocol $P_\rho$ for every partition of the inputs $z = z_1, \ldots, z_\ell$. Every protocol $P_\rho$ has to be correct on all inputs partitioned according to $\rho$.*

*The average partition communication cost of a deterministic protocol $P$ on an input $z$ is the average number of bits exchanged by $P_\rho$, where the distribution on $\rho$ is defined such that each $z_j$ is assigned as input to Alice or Bob uniformly at random. The average partition communication cost of $P$ is then the maximum over all inputs.*

*The deterministic average partition communication complexity of a function $f$ is the minimum average partition communication cost over all protocols for $f$.*

*For randomized protocols the following correctness criteria must hold: for an input $z$ the probability of $P$ giving the correct output must be at least $2/3$, over both*

*the random choices inside the $P_\rho$ and the choice of $P_\rho$ itself.*

We note that in the definition above, the notion of average partition is slightly different from the notion of random-partition we define in Section 3.1 in that we are interested in *expected* communication complexity for the case of average partition, and *worst-case* communication complexity for the case of random partition. Since a protocol with worst-case communication complexity $c$ also gives an average communication complexity $c$, a lower bound for the average partition model is also a lower bound for the random partition model. In this section, we will prove a lower bound for the average partition model.

We will show a lower bound on the randomized average partition communication complexity of the Disjointness problem disj. This complexity measure is defined as a maximum over all inputs (of the average cost of randomized protocols under a random partition). As usual we can instead (as in the easy direction of the Yao principle) bound the expected complexity of *deterministic protocols* over a hard distribution on inputs (and over the random partitions).

Recall that the partitions $\rho$ of $\{z_1, \ldots, z_\ell\}$ are chosen uniformly. Furthermore, inputs $z$ will be chosen from a distribution $\mu$ defined by Razborov. It is important that inputs and their partition are chosen independent of each other. Due to independence, when analyzing the error of a protocol, we may also choose the partition first, and then consider the error for random inputs given a fixed partition.

THEOREM A.1. *The randomized average partition communication complexity of the Disjointness problem* disj *is $\Omega(N)$.*

The theorem also holds, if Alice also gets to know all of $x$ and Bob gets to know all of $y$.

**Proof of Theorem A.1** The inputs to Disjointness are $z = x_1, \ldots, x_N, y_1, \ldots, y_N$. We first take a look at a simple property that most partitions have, namely that on most partitions many pairs $x_j, y_j$ are distributed among Alice and Bob. For simplicity we only care about pairs where $x_j$ is with Alice and $y_j$ is with Bob.

After that we follow Razborov's proof for the hardness of Disjointness under a fixed distribution $\mu$, but we have to be more flexible with parameters, and restrict attention to a part of the distribution $\mu$ that is still hard (for a given partition of the inputs). Furthermore we have to analyze certain parts of rectangles separately to keep the product properties in Razborov's proof intact.

In the following $\delta$ will be a small all-purpose constant. Denote the expected error of the protocol by $\epsilon$. This error is the expectation over inputs $z$ (from $\mu$) and over partitions $\rho$ of the input positions $\{1, \ldots, 2N\}$. We denote the expected error under a fixed partition $\rho$, i.e., the expected error of $P_\rho$, by $\epsilon_\rho$. The probability (over $\rho$) that $\epsilon_\rho$ is larger than $\epsilon/\delta$ is at most $\delta$.

Besides the error also the communication cost on input $z$ is an expectation, this time over the choice of the partition. Recall that we are dealing with deterministic protocols $P_\rho$. Denote by $c$ the expected communication over the choice of $\rho$ and the choice of $z$, and by $c_\rho$ the expectation over $z$ with $\rho$ fixed.

In our main lemma we will show that for most partitions $\rho$ this is impossible unless $c = \Omega(N)$.

We first define the input distribution used by Razborov for disj on inputs of size $N$.

DEFINITION 3. *Set $N = 4m-1$, the size of the universe of elements. To choose inputs according to Razborov's distribution one first chooses a frame. The frame consists of a partition of $\{1, \ldots, N\}$ into 3 disjoint subsets $z_x, z_y, \{i\}$ such that $|z_x| = |z_y| = 2m - 1$.*

*For the distribution $\mu_1$ on 1-inputs to* disj *one then chooses a subset $x$ of size $m$ of $z_x$ as an input to Alice, and a subset of size $m$ of $z_y$ as an input to Bob.*

*For the distribution $\mu_0$ on 0-inputs one chooses a subset $x$ of size $m - 1$ from $z_x$ and lets Alice's input be $x \cup \{i\}$, and correspondingly for Bob.*

*The distribution $\mu$ is the mixture $3/4 \cdot \mu_0 + 1/4 \cdot \mu_1$.*

Note that $\mu$ is the uniform distribution on all inputs for which both sets are of size $m$ and the intersection size is at most 1.

In our setting the partition $\rho$ of input positions (not to be confused with the partitions from the definition of $\mu$) is chosen independent of the inputs (which are chosen from $\mu$). We next observe that usually enough inputs $x_i, y_i$ are distributed among Alice and Bob.

LEMMA A.1. *Let $\rho$ be a uniformly random partition of the input variables $x_1, \ldots, x_N, y_1, \ldots, y_N$. Then with probability $1 - \delta$ we have that at least $N/4 - \sqrt{N/\delta}$ pairs $x_j, y_j$ are distributed such that Alice holds $x_i$ and Bob holds $y_i$.*

PROOF: For each pair $x_i$ is on Alice's side and $y_i$ on Bob's side with probability $1/4$. Let $C$ denote number of $i$'s for which this is the case. The expectation of $C$ is $N/4$ and the variance is $3N/16$. Hence, by Chebyshev's inequality the probability that $|C - N/4|$ is larger than $(1/\sqrt{\delta})(\sqrt{3}/4)\sqrt{N}$ is at most $\delta$. □

We call partitions $\rho$ under which more than $n = N/5$ positions $j$ are such that $x_j$ is with Alice and $y_j$ with Bob *good partitions*. In the following we analyze protocols under an arbitrary fixed good partition, showing that protocols with low communication cost will have large error. Since the expected communication is at least $1 - \delta$ times the cost under the best good partition, a linear lower bound will follow for a random partition. On the other hand, at most a $\delta$ fraction of all partitions is allowed to have larger error than $\epsilon/\delta$.

So fix any good partition $\rho$. For notational simplicity assume that $x_1, \ldots, x_n$ belong to Alice, and $y_1, \ldots, y_n$ to Bob. The remaining inputs are distributed in any other way. We consider inputs drawn from the distribution $\mu$.

Note first, that the probability (under $\mu$) that for the "intersection position" $i$ is larger than $n$, i.e., that both $x_i$ and $y_i$ might belong to Alice (or Bob) is 4/5, and in this case we assume the protocol already knows the answer without communication. So we have to assume that $\epsilon < 1/20$ to get a lower bound, since $i$ is contained in both $x$ and $y$ with probability 1/4, and hence with error 1/20 the problem to be solved is trivial. If the error is smaller than, say 1/50, then the protocol must give the correct output for most inputs with the intersecting position $i$ smaller than $n+1$.

Define $\mu_\rho$ as the distribution $\mu$ restricted to $i$ (as part of the frame) being chosen from 1 to $n$. Note that $\mu = 1/5 \cdot \mu_\rho + 4/5 \cdot \sigma$ for some distribution $\sigma$. Since the protocols $P_\rho$ we consider have error $\epsilon_\rho$ on $\mu$ (using partition $\rho$), they can have error at most $5\epsilon_\rho$ on $\mu_\rho$. So we restrict our attention to $\mu_\rho$.

$\mu_{\rho,1}$ is the distribution $\mu_1$ restricted to $i$ being from 1 to $n$, and $\mu_{\rho,0}$ is the distribution $\mu_0$ under the same restriction. We will denote the random variables of inputs chosen from $\mu_{\rho,1}$ by $\mathbf{x_0}, \mathbf{y_0}$ (indicating intersection size 0), and the random variables of inputs chosen from $\mu_{\rho,0}$ by $\mathbf{x_1}, \mathbf{y_1}$. $\mathbf{x}, \mathbf{y}$ is the pair of random variables drawn from $\mu$. Note that $\mathbf{x}$ is the random variable corresponding to the marginal of $\mu$ on the $x$-variables.

Furthermore we will denote by $\mathbf{a}, \mathbf{b}$ etc. the random variables $\mathbf{x}, \mathbf{y}$ grouped into Alice's and Bob's inputs. Note that the first $n$ variables in $\mathbf{x}$ belong to Alice and the first $n$ variables in $\mathbf{y}$ belong to Bob, the other variables are assigned in any way. Subscripts again indicate drawing inputs from $\mu_{\rho,0}$ or $\mu_{\rho,1}$.

We can now state the Main Lemma.

LEMMA A.2. *Let $\rho$ be a good partition. $\mathbf{x}, \mathbf{y}$ are chosen from $\mu_\rho$ (i.e., $\mu$ with $i < n + 1$), and partitioned into random variables as described above. Let $R = \bar{A} \times \bar{B}$ denote a rectangle of size $2^{-\zeta N}$ (under $\mu_\rho$) in the communication matrix induced by disj and $\rho$, for some small constant $\zeta > 0$.*

*Then*
$$Pr((\mathbf{a_1}, \mathbf{a_1}) \in R) \geq \alpha \cdot Pr((\mathbf{a_0}, \mathbf{a_0}) \in R),$$

*for some constant $\alpha > 0$.*

In other words, all rectangles $R$ that are trying to cover mostly 1-inputs to disj are either small or corrupted by a constant fraction of 0-inputs. A low error cover made up of rectangles hence needs to contain many rectangles.

Now we show that the main lemma implies the lower bound for disj. The randomized average partition communication complexity is lower bounded by $E_\rho E_\mu c_{\rho,z}(f)$, where $c_{\rho,z}$ is the communication cost on input $z$ under partition $\rho$. Since all good partitions together have probability $1 - \delta$, we just need to show that for any good partition $\rho$ the average communication $E_\mu c_{\rho,z} \geq \Omega(N)$.

We have $Pr_{\rho,\mu}(c_{\rho,z} > \delta c) < \delta$. Hence, if we stop all protocols $P_\rho$ once their communication on an input $z$ exceeds $c/\delta$, we increase the error by at most $\delta$. Hence in the following we can assume that for each $\rho$ and $z$ we have $c_{\rho,z} \leq c/\delta$, and in particular, that $c_\rho \leq c/\delta$.

The protocol $P_\rho$ has communication at most $c/\delta$, and corresponds to a partition of the communication matrix according to $\rho$ into $2^{c/\delta}$ rectangles with total error $\epsilon_\rho + \delta$ under $\mu$. Since $\mu_\rho$ makes up one fifth of $\mu$, the error under $\mu_\rho$ can be at most $5(\epsilon_\rho + \delta)$. Furthermore, the inputs in the support of $\mu_\rho$ (i.e., the inputs with intersection size at most 1 and intersections happening at position $n$ or less) are partitioned into $2^{c/\delta}$ rectangles with error $10(\epsilon_\rho + \delta)$. Since $\mu_\rho$ puts weight 3/4 on inputs $x, y$ that are disjoint, there must be a single rectangle $R$ that covers $(3/12) \cdot 2^{-c/\delta}$ of 1-inputs and at the same time has error at most $30(\epsilon/\delta + \delta)$ (both under $\mu_\rho$). The Main Lemma now leads to a contradiction for $\zeta N < c/\delta$ and $30(\epsilon/\delta + \delta)$ being a suitably small constant.

We now proceed to show the main lemma. In Razborov's proof the rectangles $R$ to be analyzed are product sets with respect to the divide between $x$ and $y$. For us this is not the case, because Alice and Bob hold both $x$- and $y$-inputs. However, Alice holds $x_1 \ldots, x_n$ and Bob $y_1, \ldots, y_n$. Call all the remaining variables (that are distributed arbitrarily) $O$.

In the statement of the Main Lemma $\alpha/(1 + \alpha)$ is (a lower bound on) the error of $R$ under $\mu_\rho$. Since $\alpha$ is a constant we will determine later it is enough to show a constant lower bound on the error of any large rectangle $R$. Denote the error of $R$ under $\mu_\rho$ by $\alpha'$.

The rectangle $R$ can be partitioned into smaller rectangles by fixing the variables $O$ to some value $o$. We denote a resulting rectangle by $R^o$ (note that $R^o$ is indeed a rectangle). Let $\gamma_o$ denote the average error of $R^o$ under $\mu_\rho$, normalized to $R^o$ (where $o$ is chosen according to the size of $R^o$). Clearly the expected $\gamma_o$ is $\alpha'$.

$R^o = \bar{A}^o \times \bar{B}^o$ corresponds to a rectangle in $\{0,1\}^n \times \{0,1\}^n$ by ignoring fixed input variables. $R$ itself is a rectangle in $\{0,1\}^N \times \{0,1\}^N$. $o$ fixes $x,y$ on the last $N-n$ input positions, and if we define $a$ as $x \cap \{1,\ldots,n\}$ and $b = y \cap \{1,\ldots,n\}$, then $m_a = |a|$ and $m_b = |b|$ are fixed on $R^o$.

For every $a,b$ with fixed $m_a, m_b$ there are

$$f_{a,b} = \binom{N-n}{m-m_a} \cdot \binom{N-n-m+m_a}{m_b}$$

extensions $o$. And hence as many (disjoint) rectangles $R^o$ in $R$. Later we will restrict our attention to $o$ for which $m_a, m_b$ are very close to $n/4$. Recall that $R$ has size $\mu_\rho(R) \geq 2^{-\zeta N}$ by the assumptions in the Main Lemma.

To apply a modification of Razborov's original argument for disj on $\{0,1\}^n \times \{0,1\}^n$ we want to find an $o$ such that $R^o$

1. is large (as a rectangle in $\{0,1\}^n \times \{0,1\}^n$)

2. has small error

3. the size of $x \cap \{1, \ldots n\}$ and $y \cap \{1, \ldots, n\}$ is roughly $n/4$ for all inputs in $R^o$.

The rectangles $R^o$ with $\mu_\rho(R^0) < f_{a,b}^{-1} \cdot 2^{-2\zeta N}$ contribute at most $2^{-2\zeta N}$ to $R$ and can thus be disregarded without changing the error much (even if the small $R^o$ contain only 1-inputs). We choose $o$ now by picking $x,y$ from $\mu_\rho$ restricted to $R$, and removing $x_1, \ldots, x_n, y_1, \ldots, y_n$. This clearly chooses a $R^o$ by size, and hence the expected error is at most $2\alpha'$, and the probability that the error is larger than $2\alpha'/\delta$ is at most $\delta$.

We now have to consider the size of sets $a = x \cap \{1, \ldots, n\}$ and $b = y \cap \{1, \ldots, n\}$. Note that $o$ fixes some $k_x$ of the $x$-variables to 1 and some $k_y$ of the $y$-variables, and the size of $a$ is $m - k_x$, the size of $b$ is $m - k_y$.

The following claim follows from a simple application of the Chernoff inequality.

CLAIM 1. *Choose $o$ by choosing $x,y$ from $\mu_\rho$ and discarding the variables outside $O$.*

$$Pr(|\ |a| - n/4\ | > \kappa n\ or\ |\ |b| - n/4\ | > \kappa n) \leq 2^{-\kappa^2 n/3}.$$

For a small enough $\kappa$ (namely $\kappa^2/3 \cdot n \ll \zeta N$) this means that when we choose $o$ as above (i.e., $x,y$ from $R$ according to $\mu$ and dropping the unneeded variables), then with probability $> 2/3$ the size of $a$ and $b$ are within $\pm \kappa n$ of $n/4$.

Hence we can find $o$ such that the above 3 criteria are satisfied, and all that is left to do is to show that $R_o$ must have large error after all. By now we are almost in the same situation as Razborov, except that the sets $a,b$ are not exactly of size $n/4$.

Denote by $m_a$ and $m_b$ the sizes of $a,b$ respectively. We consider the distribution $\nu$ which is defined as follows.

DEFINITION 4. $n = 4m' - 1$. *To choose inputs one first chooses a frame. There are $n$ input positions. The frame consists of a partition of $\{1, \ldots, n\}$ into 3 disjoint subsets $z_a, z_b, \{i\}$ such that $|z_a| = |z_b| = 2m' - 1$.*

*For the distribution $\nu_1$ on 1-inputs to disj one then chooses a subset $a$ of size $m_a$ of $z_a$ as an input to Alice, and a subset of size $m_b$ of $z_b$ as an input to Bob.*

*For the distribution $\mu_0$ on 0-inputs one chooses a subset $a$ of size $m_a - 1$ from $z_a$ and lets Alice's input be $a \cup \{i\}$, and correspondingly for Bob.*

*The distribution $\nu$ is the mixture $3/4 \cdot \nu_0 + 1/4 \cdot \nu_1$.*

Note that $\nu$ can be obtained from $\mu_\rho$ by dropping the last $N - n$ elements of the universe. Since $\mu$ is invariant under permuting the last $N - n$ or the first $n$ elements of the universe, for all $x,y$ and $a,b$ obtained from them $\nu(a,b) = \mu_\rho(x,y) \cdot f_{a,b}$. In our case all $f_{a,b}$ are within $2^{O(\kappa n)}$ of each other. This means that our $R^o$ has size $\nu(R^o) \geq 2^{-2\zeta N - O(\kappa N)}$.

What is left to do is to prove Razborov's Main Lemma under the modification that the sets $x,y$ have fixed sizes that are slightly different from $n/4$. The argument is almost entirely the same, with some estimates looser than in the original.

We now restate that Lemma. In this restatement we re-use the notation $\mathbf{a_1}$ etc., which now refer to the domain $\{0,1\}^n \times \{0,1\}^n$ and $\nu$ etc.

LEMMA A.3. *$a,b$ are chosen from $\nu$ (as above). Let $R = \bar{A} \times \bar{B}$ denote a rectangle of size $2^{-\zeta'}$, for some small constant $\zeta' > 0$.*

*Then*

$$Pr((\mathbf{a_1}, \mathbf{a_1}) \in R) \geq \alpha \cdot Pr((\mathbf{a_0}, \mathbf{a_0}) \in R),$$

*for some constant $\alpha > 0$.*

We omit the details of this. As discussed above this implies our Main Lemma, and hence the lower bound for disj.

To see that the above proof also applies if Alice knows all of $x$ and Bob knows all of $y$ note that we actually analyze the rectangles $R^0$, in which the extra inputs are fixed anyway.

## B  Proof of Lemma 3.1

We will use the following lower bound which can be derived using basic facts in communication complexity. We sketch its proof for completeness.

THEOREM B.1. *For some $\epsilon > 0$, $R_{0,\epsilon}^{cc-pub}(\text{eq}) \geq b$.*

*Proof.* [Proof sketch] We need the following additional notations. Let $N^1(f)$ denote the *nondeterminic communication complexity* of computing $f$ (see, e.g., [22, Definition 2.3]). It is well-known that $N^1(\text{eq}) \geq b$ (see, e.g. [22, Example 2.5]). Note further that $R_{0,\epsilon}^{cc-pub}(f) \geq N^1(f)$ (see, e.g., [22, Proposition 3.7][6]). Thus, $R_{0,\epsilon}^{cc-pub}(\text{eq}) \geq N^1(\text{eq}) \geq b$ as claimed.

*Proof.* [Proof Lemma 3.1] We will now prove that if $R_{0,\epsilon}^{rcc-pub}(\text{eq}) < b/4 - \sqrt{6b \ln b}$, then $R_{0,\epsilon}^{cc-pub}(\text{eq}) < b$, thus contradicting Theorem B.1. This immediately implies that $R_{0,\epsilon}^{rcc-pub}(\text{eq}) = \Omega(b)$, as claimed in Lemma 3.1. Let $\mathcal{A}$ be a protocol in the random-partition model which requires strictly less than $b/4$ bits of communication. Alice and Bob simulates $\mathcal{A}$ as follows. First, they use shared randomness to pick a random partition of $x$ and $y$, denoted by $\mathcal{P}$. Let $S_B$ be the set of positions in $x$ that Bob receives in partition $\mathcal{P}$. If $|S_B| > b/2 + (1/2)\sqrt{6b \ln b}$, then they stop the simulation and answer 0 (representing "$x \neq y$"). Note that this could cause an error when $x = y$; however, by Chernoff's bound (e.g. [30, Section 4.2.2]), this happens with probability at most $2/b$. If Alice and Bob do not stop the simulation, Alice sends to Bob bits in $x$ at positions in $S_B$. This causes a communication of $b/2 + (1/2)\sqrt{6b \ln b}$ (note that Alice simply sends $|S_B|$ bits in order and Bob can figure out which bit belongs to which position since he also knows the partition $\mathcal{P}$). Next, Bob checks whether bits in $x$ sent from Alice is the same as his bits (in the same position). If not, he immediately answers 0 ("$x \neq y$"); otherwise, we let $S_A$ be the set of positions *not* in $S_B$ such that Alice receives bits in $y$ at positions in $S_A$. Bob checks if $|S_A| > (b - |S_B|)/2 + (1/2)\sqrt{6b \ln b}$. If this is the case, then Alice and Bob stops the simulation and answer 0. This again could cause an error but it happens with probability at most $2/b$ by Chernoff's bound. If this does not happen, Bob sends bits in $y$ at positions in $S_A$ to Alice. This causes at most $(b - |S_B|)/2 + (1/2)\sqrt{6b \ln b}$ bits of communication. Then, Alice and Bob simulates $\mathcal{A}$, which will finish after less than $b/4 - \sqrt{6b \ln b}$ bits of communication and produces an answer that is $(\epsilon, 0)$-error. Alice and Bob use this answer as their answer. Combining with the previous errors, we have that Alice and Bob's answer is $(\epsilon + 4/b, 0)$-error. (This error comes from the worst-case scenario where $\mathcal{A}$ could be always correct for the partitions that Alice and Bob stops. This gives the additional error of $4/b$.) Moreover, the number of communication bits that their simulation requires is less than $|S_b| + (b - |S_b|)/2 + (1/2)\sqrt{6b \ln b} + b/4 - \sqrt{6b \ln b}$ which is less than $b$ since $|S_b| \leq b/2 + (1/2)\sqrt{6b \ln b}$. This implies that $R_{0,\epsilon}^{cc-pub}(\text{eq}) < b$ as desired.

Note that in the reduction above Alice knows $x$ and Bob knows $y$. Thus, the claimed lower bound holds even when Alice and Bob have this information in addition to the random partition.

---

[6]We note that our notations are slightly different from [22]. It can be checked from [22, Definitions 3.1 and 3.3] that the definition of $R_{0,\epsilon}^{rcc-pub}$ is the same as $R^1$, i.e. they correspond to the case where we allow an error only when $f(x,y) = 1$.